\documentclass[twocolumn]{webofc}

\usepackage[varg]{txfonts}   
\usepackage[mathscr,scaled=1.15]{urwchancal}
\DeclareFontFamily{OT1}{pzc}{}
\DeclareFontShape{OT1}{pzc}{m}{it}%
{<-> s * [1.15] pzcmi7t}{}
\DeclareMathAlphabet{\mathpzc}{OT1}{pzc}{m}{it}

\usepackage{color}

\definecolor{prdblue}{rgb}{0.133,0.118,0.498}
\usepackage[colorlinks=true, pdfstartview=FitV, linkcolor=prdblue, citecolor= prdblue, urlcolor=prdblue]{hyperref}

\begin{document}

\begin{flushleft}
NJU-INP 007/19
\end{flushleft}
\vspace*{-2cm}

\title{Form factors for the Nucleon-to-Roper electromagnetic transition at large-$Q^2$}
%
%

\author{\firstname{José} \lastname{Rodr\'{\i}guez-Quintero}\inst{1}\fnsep\thanks{\email{jose.rodriguez@dfaie.uhu.es}} \and
        \firstname{Daniele} \lastname{Binosi}\inst{2} \and 
        \firstname{Chen} \lastname{Chen}\inst{3} \and
        \firstname{Ya} \lastname{Lu}\inst{4.5} \and 
         \firstname{Craig D.} \lastname{Roberts}\inst{4,5} \and 
		 \firstname{Jorge} \lastname{Segovia}\inst{6} 
}

\institute{Department of Integrated Sciences, University of Huelva, E-21071 Huelva, Spain
\and        
 		European Centre for Theoretical Studies in Nuclear Physics
and Related Areas (ECT$^\ast$) and Fondazione Bruno Kessler\\ Villa Tambosi, Strada delle Tabarelle 286, I-38123 Villazzano (TN) Italy
\and
Justus-Liebig-Universität Giessen, Institut f\"ur Theoretische Physik, Heinrich-Buff-Ring 16,  35392 Giessen,  Germany 
\and
School of Physics, Nanjing University, Nanjing, Jiangsu 210093, China
\and
Institute for Nonperturbative Physics, Nanjing University, Nanjing, Jiangsu 210093, China
\and
	Departamento de Sistemas F\'{\i}sicos, Qu\'{\i}micos y Naturales,
Universidad Pablo de Olavide, E-41013 Sevilla, Spain	
          }

\abstract{%
We report on a recent calculation of all Roper-related electromagnetic transtions form factors, covering the range of energies that next-to-come planned experiments are expected to map. Direct reliable calculations were performed, within a Poincar\'e covariant approach of the three-body bound-state problem, up to $Q^2/m^2_N$=6; approximated then by applying the Schlessinger point method and the results eventually extended up to $Q^2/m^2_N\simeq$12 {\it via} analytic continuation.    
}
\maketitle
\section{Introduction}
Albeit there is no doubt about the nature of nucleons as bound states made of three valence-quarks, that of their first excited states -- $N(1440)\,1/2^+$, $N(1535)\,1/2^-$ -- is less certain.  The parity-positive one, discovered in 1963 \cite{Roper:1964zza, Bareyre:1964pcg, Auvil:1964qxg, Adelman:1964jmp, Roper:1965pfb} and dubbed ``Roper resonance'' since then, puzzled immediately the nuclear physicists because, \emph{e.g}.\ a second positive-parity state lying above the first negative-parity one in the baryon spectrum had been predicted by a wide array of constituent-quark potential models (\emph{e.g.}\,see \cite{Crede:2013sze}). During the last 20 years, a combined effort for the acquisition of a vast amount of high-precision proton-target exclusive electroproduction data, for their analysis with sophisticated tools from dynamical reaction theory, and for the formulation of wide-ranging application of a Poincar\'e covariant approach to the continuum bound-state problem in quantum field theory led to the wide acceptance about the Roper being the first radial excitation of the nucleon\,\cite{Burkert:2019bhp}. And about its being sensibly described as a well-defined dressed-quark core, augmented by a meson cloud which both reduces the Roper's core mass by approximately 20\% and contributes materially to the electroproduction transition form factors at low-$Q^2$. 

The high-$Q^2$ electroproduction data were crucial to reaching this understanding of the Roper and, with a new era of experiments beginning at the upgraded JLab12 in the near future, are expected to sharpen the contemporary picture for the Roper resonance by fostering a unified explanation of elastic and transtion form factors for nucleons, $\Delta$-baryon and Roper resonances. To pursue this goal, given that JLab\,12 will deliver results for the $Q^2$-dependence of $R^{0,+}$ electrocouplings, reaching to $Q^2\approx 12\,m_N^2$ ($m_N$ for the nucleon mass) \cite{Mokeev:2018zxt, Carman:2018fsn, Cole:2018faq}, we have recently reported\,\cite{Chen:2018nsg} calculations of all Roper-related transition form factors covering the $Q^2$-domain that planned experiments expect to map. We will herein sketch these calculations, grounded on the framework of many previous Poincar\'e-covariant continuum analyses of the three valence-quark bound-state problems associated with the nucleon, $\Delta$-baryon and Roper resonance\,\cite{Segovia:2014aza, Roberts:2015dea, Segovia:2015hra, Segovia:2016zyc, Chen:2017pse, Mezrag:2017znp, Roberts:2018hpf}.

%
%
%
%
%


\section{Nucleon and Roper Structure}
\label{secNRStructure}
\begin{figure}[t]
\centerline{%
\includegraphics[clip, width=0.45\textwidth]{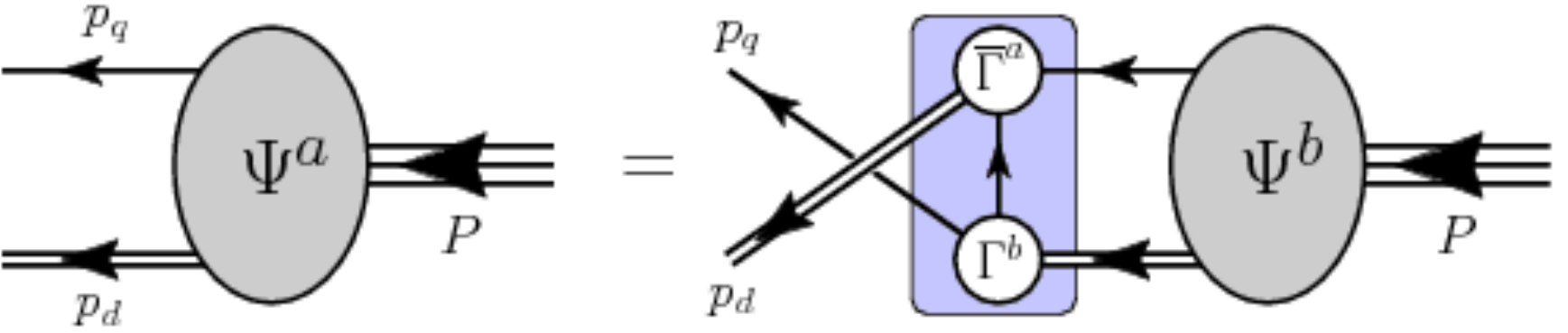}}
\caption{\label{figFaddeev}
Poincar\'e covariant Faddeev equation: a linear integral equation for the matrix-valued function $\Psi$, being the Faddeev amplitude for a baryon of total momentum $P= p_q + p_d$, which expresses the relative momentum correlation between the dressed-quarks and -diquarks within the baryon.  The shaded rectangle demarcates the kernel of the Faddeev equation: \emph{single line}, dressed-quark propagator; $\Gamma$,  diquark correlation amplitude; and \emph{double line}, diquark propagator. }
\end{figure}

The structure of baryons, treated as three--valence-body bound-states, should be revealed by the solutions of Poincar\'e-covariant Faddeev equation\,\cite{Cahill:1988dx, Burden:1988dt, Cahill:1988zi, Reinhardt:1989rw, Efimov:1990uz}, which sums all possible exchanges and interactions that can take place between the three valence dressed-quarks. Notwithstanding this, soft (nonpointlike) diquark correlations have been predicted to exist within baryons, owing to the use of a realistic quark-quark interaction \cite{Binosi:2014aea, Binosi:2016nme, Rodriguez-Quintero:2018wma} and appropriate implementation of chiral symmetry breaking (DCSB) \cite{Segovia:2015ufa}. This strongly supports a truncation of the three-body Faddeev equation such that the problem of determining the properties of the baryon's dressed-quark core transforms into that of solving the linear, homogeneous matrix equation depicted in Fig.\,\ref{figFaddeev}.

Then, following Ref.\,\cite{Segovia:2015hra}, the Faddeev equation can be solved to obtain masses and Poincar\'e-covariant wave functions for the nucleon and its first radial excitation. The masses are found to be 1.18 and 1.73 GeV, respectively, for these systems, which are predicted to be constituted only from isoscalar-scalar and isovector-pseudovector diquarks, with negligible impact from correlations in the pseudoscalar and vector channels. The masses correspond to the locations of the two lowest-magnitude $I=1/2$, $J^P=1/2^+$ poles in the three dressed-quark scattering problem, while the associated residues are the canonically-normalized Faddeev wave functions, depending upon $(\ell^2,\ell \cdot P)$, where $\ell$ is the quark-diquark relative momentum and $P$ is the baryon's total momentum.  

Then, with the wave function in hand, the zeroth Chebyshev moments of all its $S$-wave components, 
\begin{equation}
{\mathpzc W}(\ell^2;P^2) = \frac{2}{\pi} \int_{-1}^1 \! du\,\sqrt{1-u^2}\,
{\mathpzc W}(\ell^2,u; P^2)\,,
\end{equation}
where $u=\ell\cdot P/\sqrt{\ell^2 P^2}$,
can be computed and put under examination; and they can be seen to be either positive- or negative-definite for the ground-state nucleon, while a single zero is exhibitted for the first excited state (See, \emph{e.g}.\ Ref.\,\cite{Segovia:2015hra}, Fig.\,2, or Ref.\,\cite{Chen:2017pse}, Fig.\,4.), strongly indicating that this state is a radial excitation of the quark-diquark system. 
 
Concerning the masses, they can be confronted to the empirical values of the pole locations for the first two states in the nucleon channel, which are: $0.939\,$GeV for the nucleon; and two poles for the Roper, $1.357 - i \,0.076$, $1.364 - i \, 0.105\,$GeV,\,\cite{Suzuki:2009nj}. They differ at first glance, but the discrepancies can be well understood: The solutions of the Faddeev equation with the kernel in Fig.\,\ref{figFaddeev} should be interpreted as the \emph{dressed-quark core} of the bound-state, instead of representing the fully-dressed observable object, because the kernel omits\,\cite{Eichmann:2008ae, Eichmann:2008ef} all those resonant contributions which may be associated with the meson-baryon-cloud finite-state-interactions (MB FSIs) resummed in DCC models  \cite{JuliaDiaz:2007kz, Suzuki:2009nj, Kamano:2010ud, Ronchen:2012eg, Kamano:2013iva, Doring:2014qaa, Kamano:2018sfb}. Therefore, the critical comparison is not between the computed dressed-quark core masses and empirical values of the pole-positions but between the former and the values determined for the meson-undressed bare bound-state; \emph{e.g.} for the Roper, \emph{viz}.\ (in GeV):
\begin{equation}
\label{eqMassesA}
\begin{array}{l|ccc|c}
    & \mbox{R}_{{\rm core}}^{\mbox{\footnotesize \cite{Segovia:2015hra, Chen:2017pse}}}
    & \mbox{R}_{{\rm core}}^{\mbox{\footnotesize \cite{Lu:2017cln}}}
    & \mbox{R}_{{\rm core}}^{\mbox{\footnotesize \cite{Wilson:2011aa}}}
    & \mbox{R}_{\rm DCC\,bare}^{\mbox{\footnotesize \cite{Suzuki:2009nj}}} \\\hline
 \mbox{mass} & 1.73 & 1.82 & 1.72 & 1.76
\end{array}\,;
\end{equation}
The DCC bare-Roper mass agreeing remarkably well with the quark core results obtained using both a QCD-kindred interaction \cite{Segovia:2015hra} and refined treatments of a strictly-implemented vector$\,\otimes\,$vector contact-interaction \cite{Wilson:2011aa, Lu:2017cln}. An alternative test comes with clothing nucleon's dressed-quark core by including resonant contributions to the kernel. One thus produces a physical nucleon and $\Delta$-baryon masses which are $\approx 0.2$\,GeV lower than those of the core \cite{Ishii:1998tw, Hecht:2002ej,JuliaDiaz:2007kz}. Consistently, our estimate for the nucleon is 0.2\,GeV greater than the empirical value.


\section{Electromagnetic Currents}
\label{secEMcurrent}
The next brick for the desired computation of elastic and transition form factors is the electromagnetic current carrying the interaction between the participating states. When the initial and final states are $I=1/2$, $J=1/2^+$ baryons, that current is completely specified by two form factors, \emph{viz}.
\begin{equation}
\bar u_{f}(P_f)\big[ \gamma_\mu^T F_{1}^{fi}(Q^2)+\frac{1}{m_{{fi}}} \sigma_{\mu\nu} Q_\nu F_{2}^{fi}(Q^2)\big] u_{i}(P_i)\,,
\label{NRcurrents}
\end{equation}
where: $u_{i}$, $\bar u_{f}$ are, respectively, Dirac spinors describing the incoming/outgoing baryons, with four-momenta $P_{i,f}$ and masses $m_{i,f}$ so that $P_{i,f}^2=-m_{i,f}^2$; $Q=P_f-P_i$; $m_{{fi}} = (m_f+m_{i})$; and $\gamma^T \cdot Q= 0$.

\begin{figure}[!t]
\centerline{\includegraphics[clip,width=1.0\linewidth]{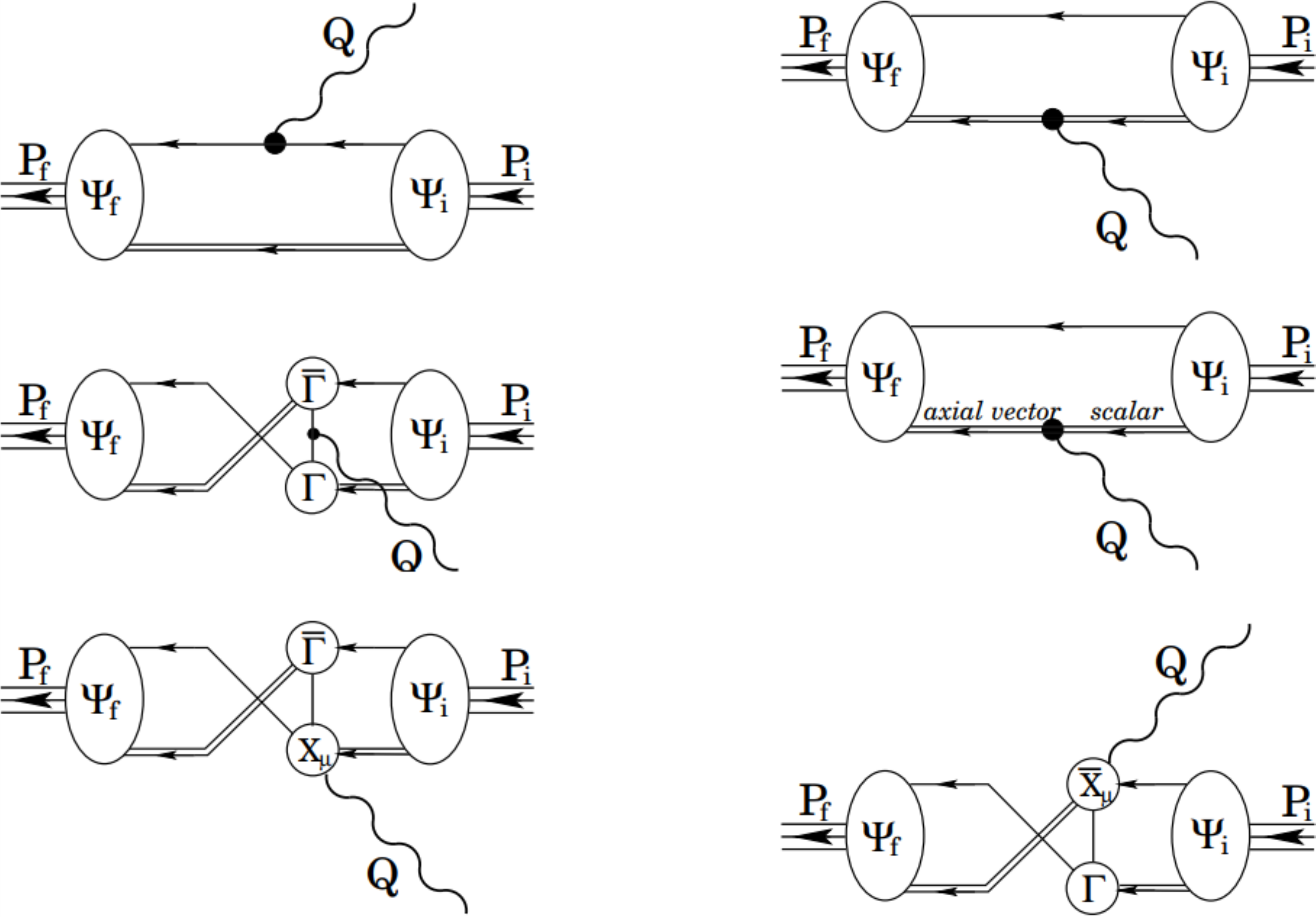}}
\caption{\label{vertexB}
Vertex that ensures a conserved current for on-shell baryons that are described by the Faddeev amplitudes produced by the equation depicted in Fig.\,\ref{figFaddeev}: \emph{single line}, dressed-quark propagator; \emph{undulating line}, photon; $\Gamma$,  diquark correlation amplitude; and \emph{double line}, diquark propagator.  Diagram~1 is the top-left image; the top-right is Diagram~2;
and so on, with Diagram~6 being the bottom-right image.
(Details are provided in Ref.\,\cite{Segovia:2014aza},  Appendix~C.)
}
\end{figure}

\begin{figure*}[t!]
\begin{center}
\begin{tabular}{lr}
\includegraphics[clip,width=0.38\linewidth]{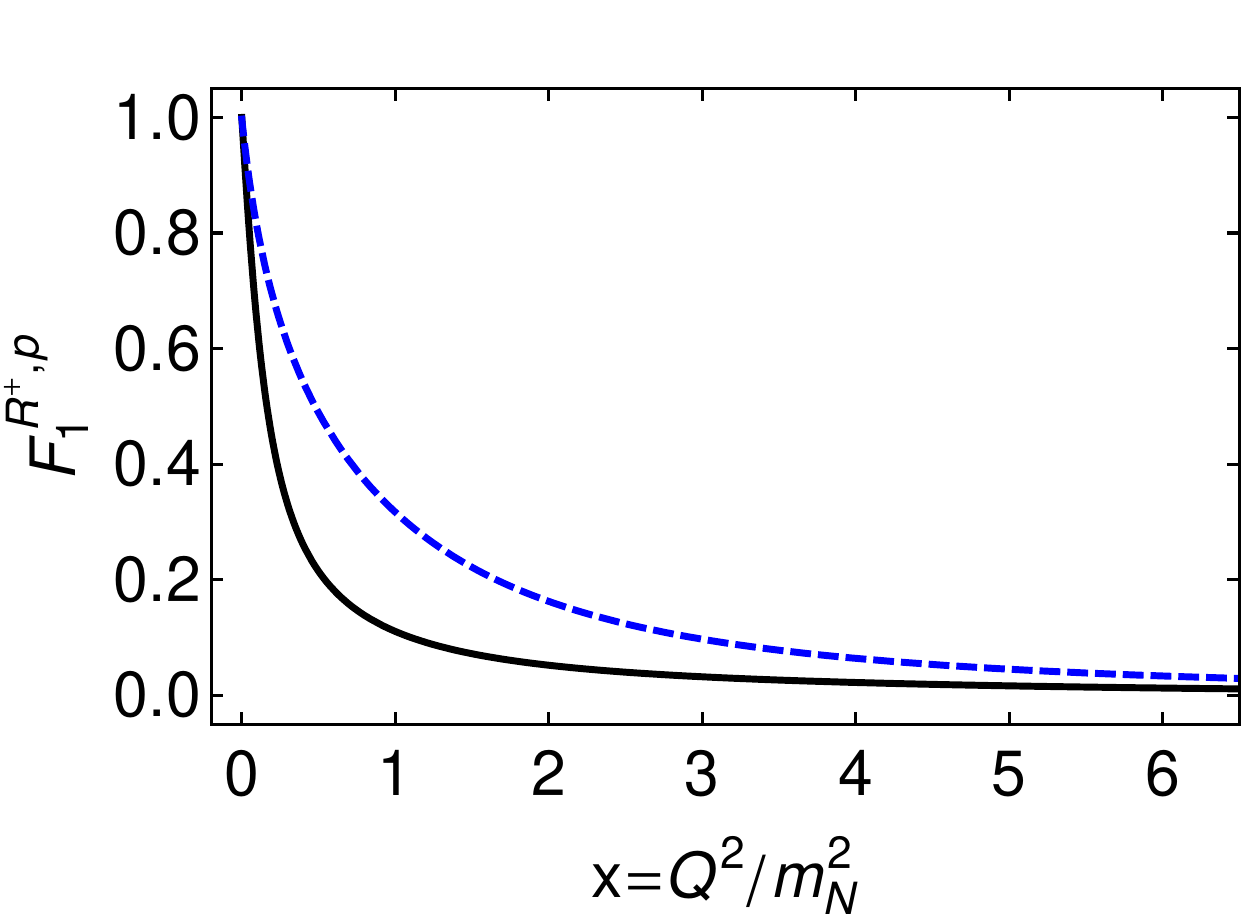}&
\includegraphics[clip,width=0.40\linewidth]{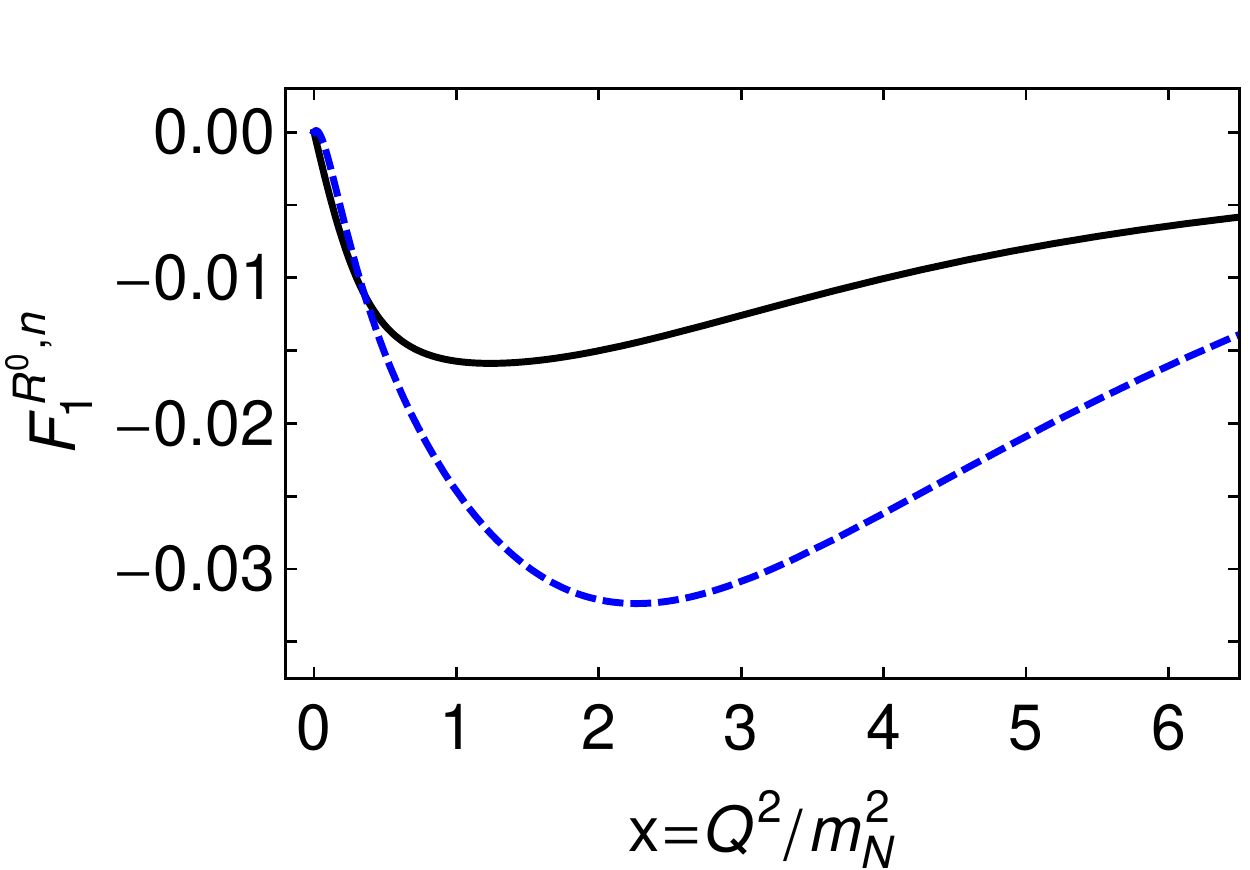}
\vspace*{-1.05cm}
\\ 
\vspace*{-0.5cm}

%
\includegraphics[clip,width=0.38\linewidth]{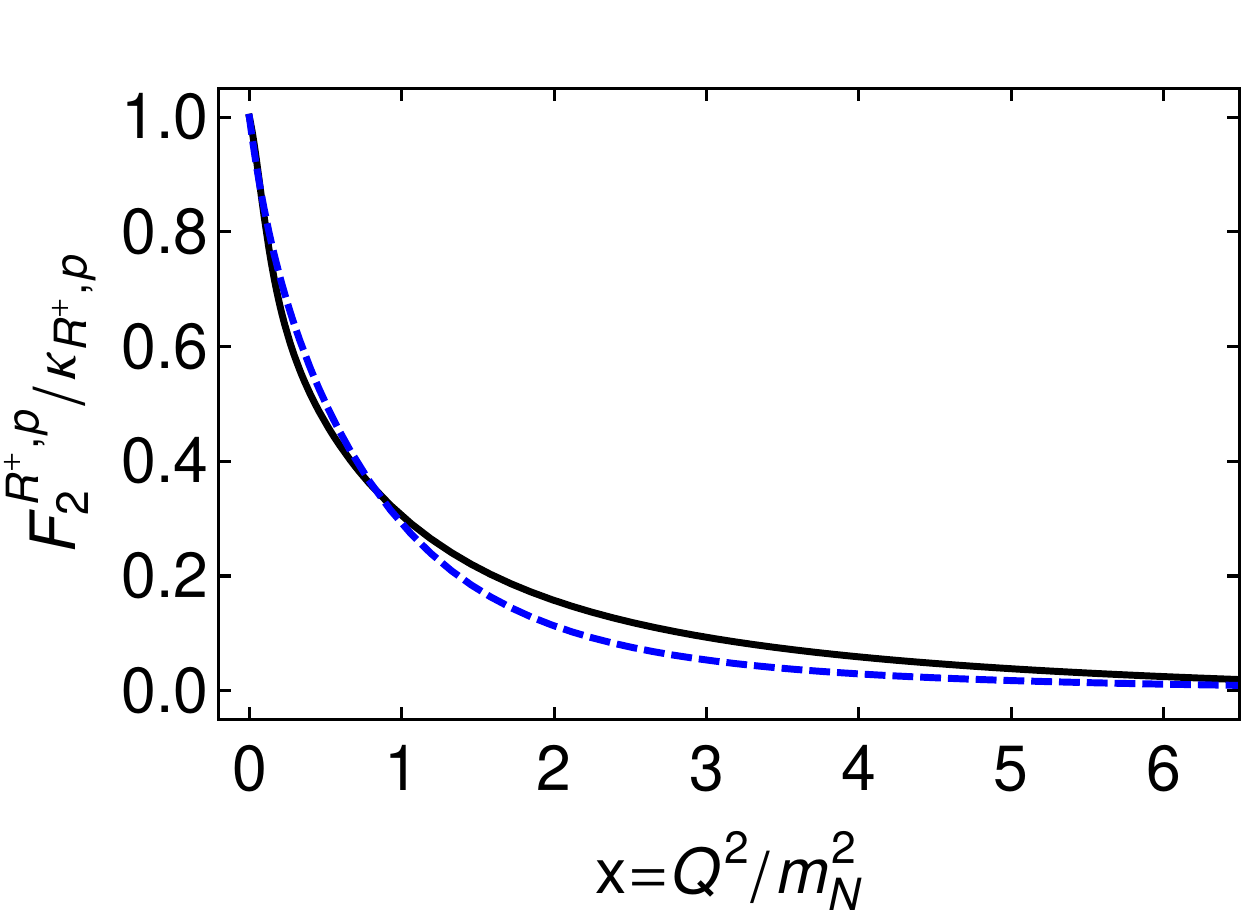} &
\includegraphics[clip,width=0.38\linewidth]{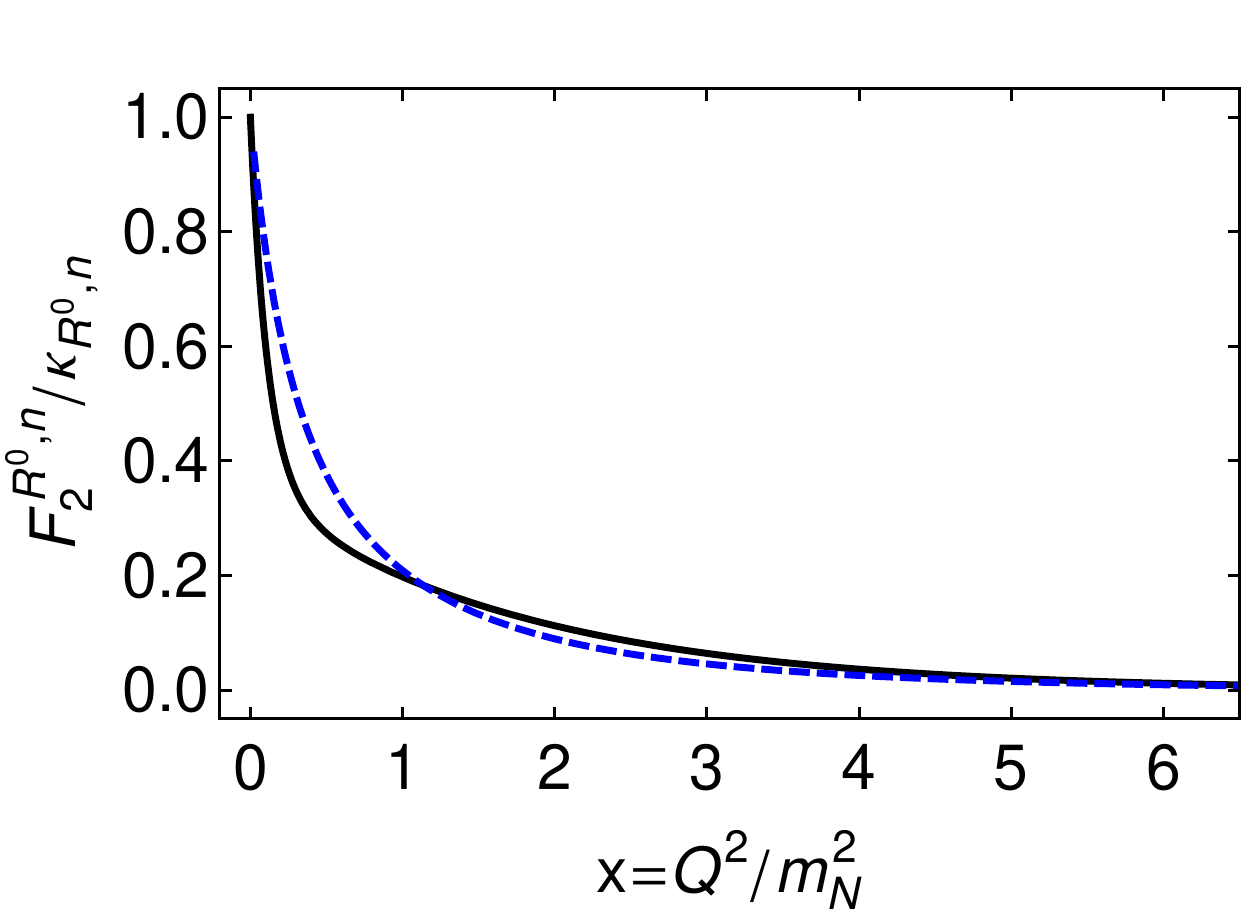}
\end{tabular}
\end{center}
\caption{\label{elastic}
Solid (black) curves -- Dirac (upper panels) and Pauli (lower) elastic electromagnetic form factors associated with the dressed-quark cores of the charged (left) and neutral (right) Roper systems.
Dashed (blue) curves -- analogous results for proton and neutron.
($\kappa_{R,N} = F_2^{R,N}(x=0)$; $x=Q^2/m_N^2$, where $M_N = 1.18\,$GeV is the nucleon' dressed-quark core mass.)
}
\end{figure*}

The interaction of a photon with a baryon generated by the Faddeev equation in Fig.\,\ref{figFaddeev} can be sufficiently expressed by a vertex including the six terms depicted in  Fig.\,\ref{vertexB}\,\cite{Oettel:1999gc, Segovia:2014aza}, with the photon separately probing the quarks and diquarks in various ways. Then, elastic and transition electromagnetic form factors involving the nucleon and Roper may be dissected in two separate ways, each of which can be considered as a sum of three distinct terms, namely: 

\noindent
{\bf DD=diquark dissection}: [\emph{DD1}] sca\-lar diquark, $[ud]$, in both the initial- and final-state baryon; 
[\emph{DD2}] pseudovector diquark, $\{qq\}$, in both the initial- and final-state baryon; 
and [\emph{DD3}] a different diquark in the initial- and final-state baryon.

\noindent 
{\bf DS = scatterer dissection}: [\emph{DS1}] 
photon strikes a bystander dressed-quark (Diagram~1 in Fig.\,\ref{vertexB});
[\emph{DS2}] photon interacts with a diquark, elastically or causing a transition scalar\,$\leftrightarrow$\,pseudovector (Diagrams~2 and 4 in Fig.\,\ref{vertexB}); 
and [\emph{DS3}] photon strikes a dressed-quark in-flight, as one diquark breaks up and another is formed (Diagram ~3 in Fig.\,\ref{vertexB}), or appears in one of the two associated ``seagull'' terms (Diagrams~5 and 6).

\begin{figure*}[t!]
\begin{center}
\begin{tabular}{lr}
\includegraphics[clip,width=0.38\linewidth]{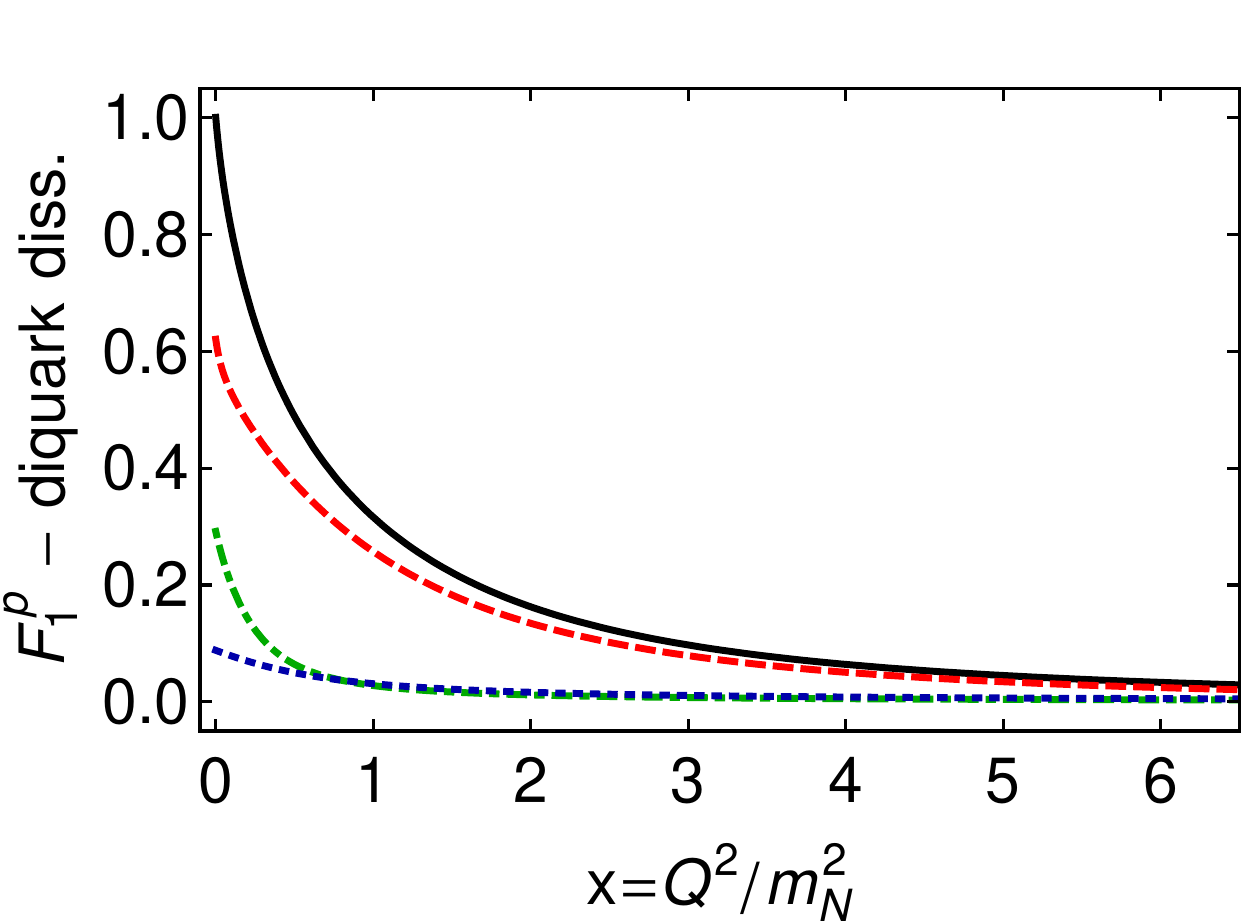}&
\includegraphics[clip,width=0.38\linewidth]{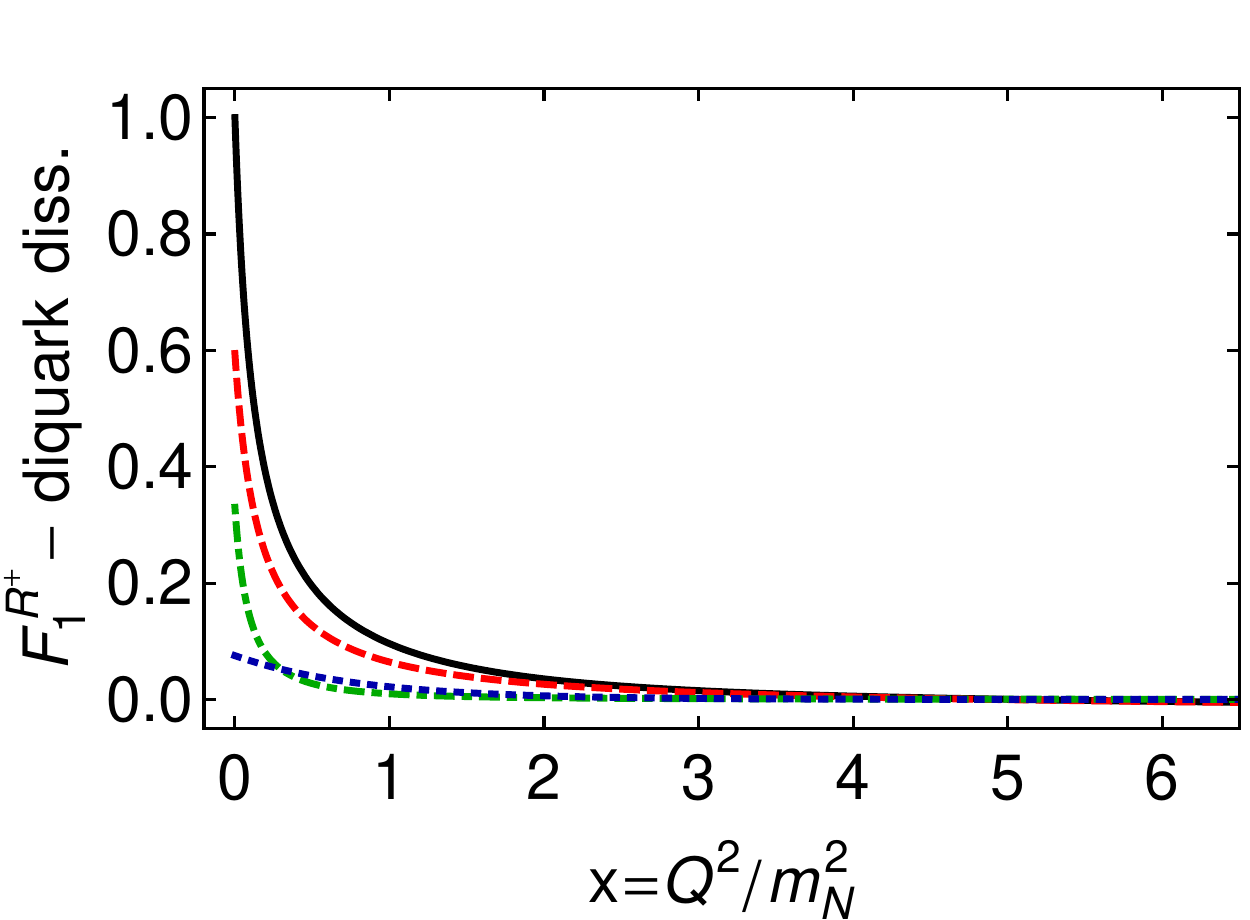}
\vspace*{-1.05cm}
\\
\vspace*{-0.5cm}
%
\includegraphics[clip,width=0.38\linewidth]{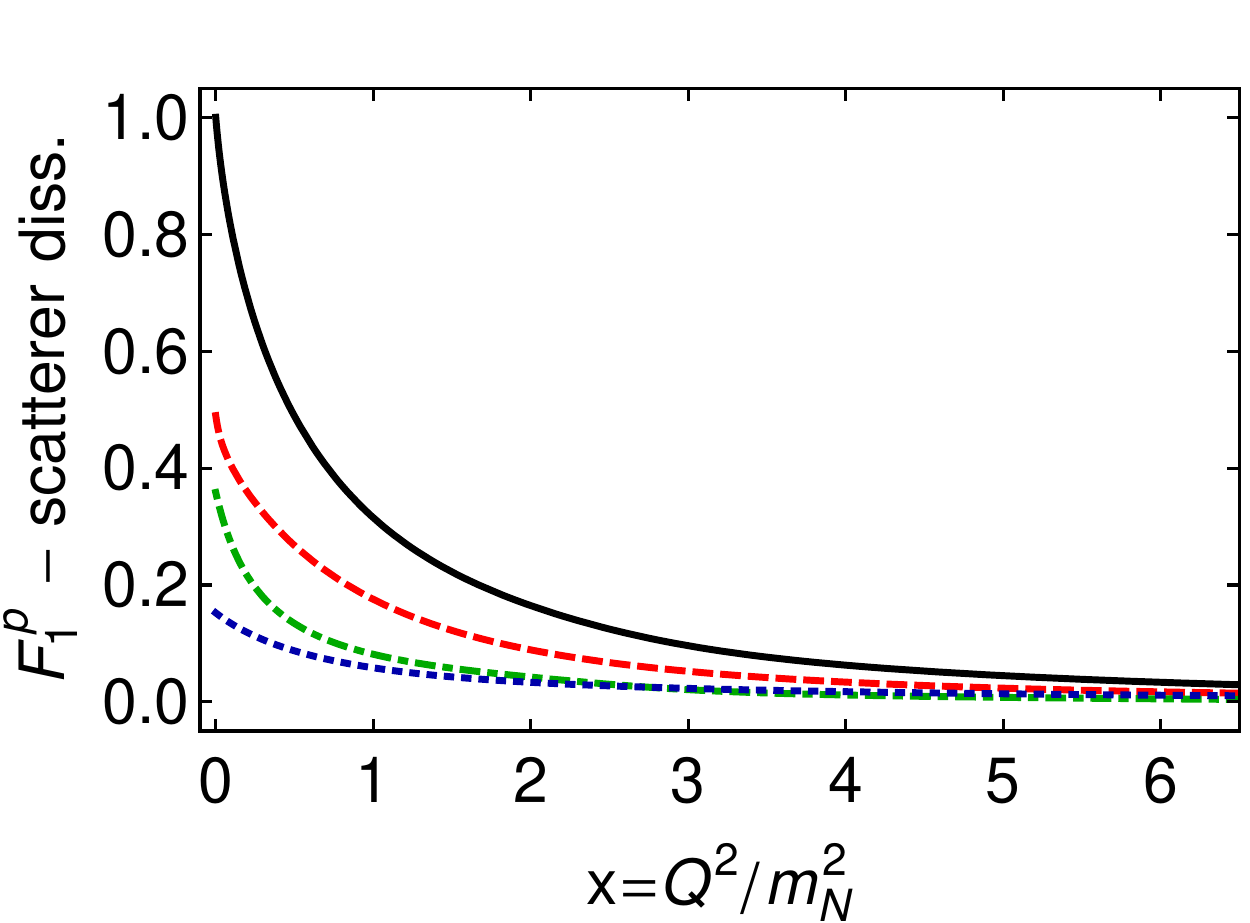} &
\includegraphics[clip,width=0.38\linewidth]{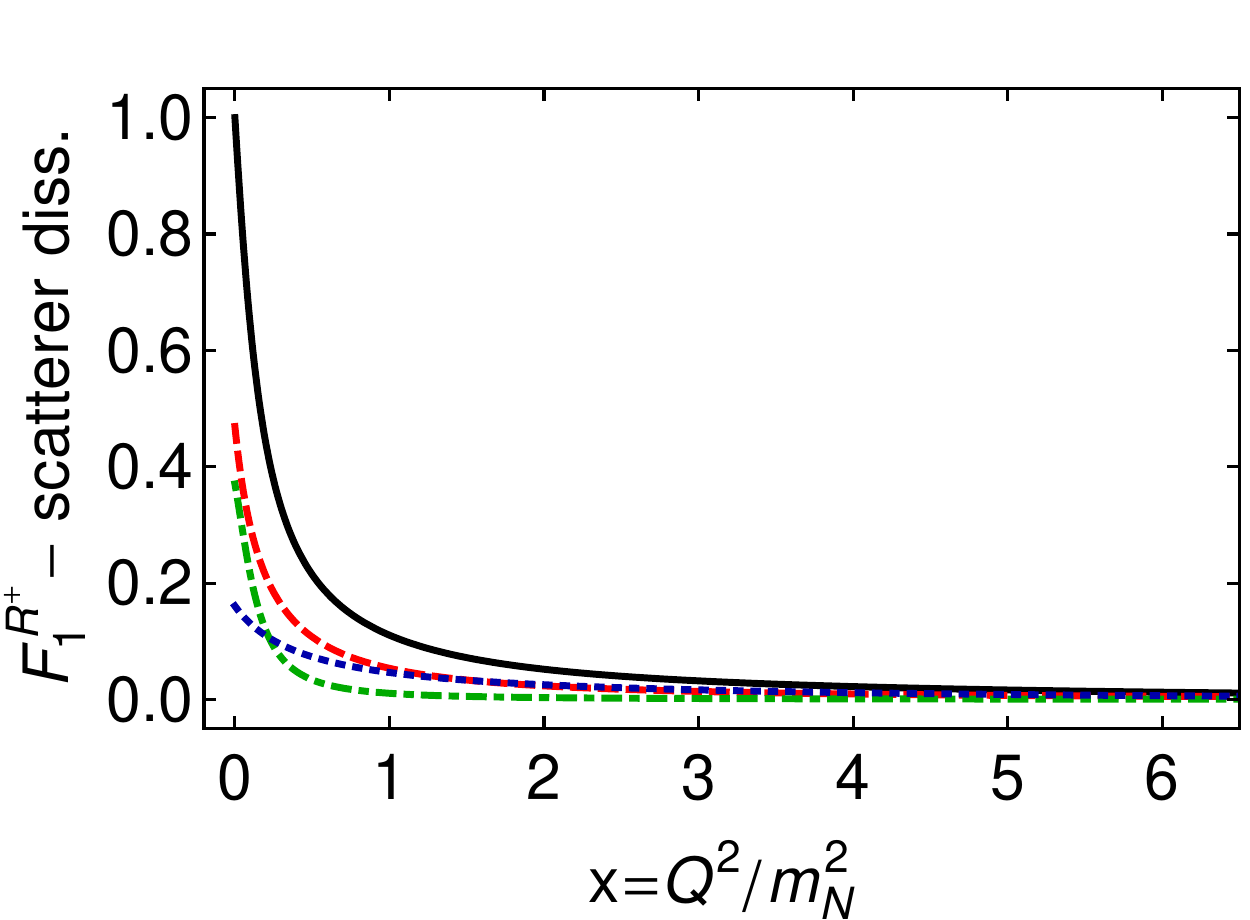}
\end{tabular}
\end{center}
\caption{\label{elasticdissectF1}
Dirac form factors of the proton (left) and charged-Roper (right).
\emph{Upper panels} -- diquark breakdown: \emph{DD1} (dashed red), scalar diquark in initial and final baryon; \emph{DD2} (dot-dashed green), pseudovector diquark in both initial and final states; \emph{DD3} (dotted blue), scalar diquark in incoming baryon, pseudovector diquark in outgoing baryon, and vice versa.
\emph{Lower panels} -- scatterer breakdown: \emph{DS1} (red dashed), photon strikes an uncorrelated dressed quark; \emph{DS2} (dot-dashed green), photon strikes a diquark; and \emph{DS3} (dotted blue), diquark breakup contributions, including photon striking exchanged dressed-quark.
}
\end{figure*}

\begin{figure*}[!t]
\begin{center}
\begin{tabular}{lr}
\includegraphics[clip,width=0.38\linewidth]{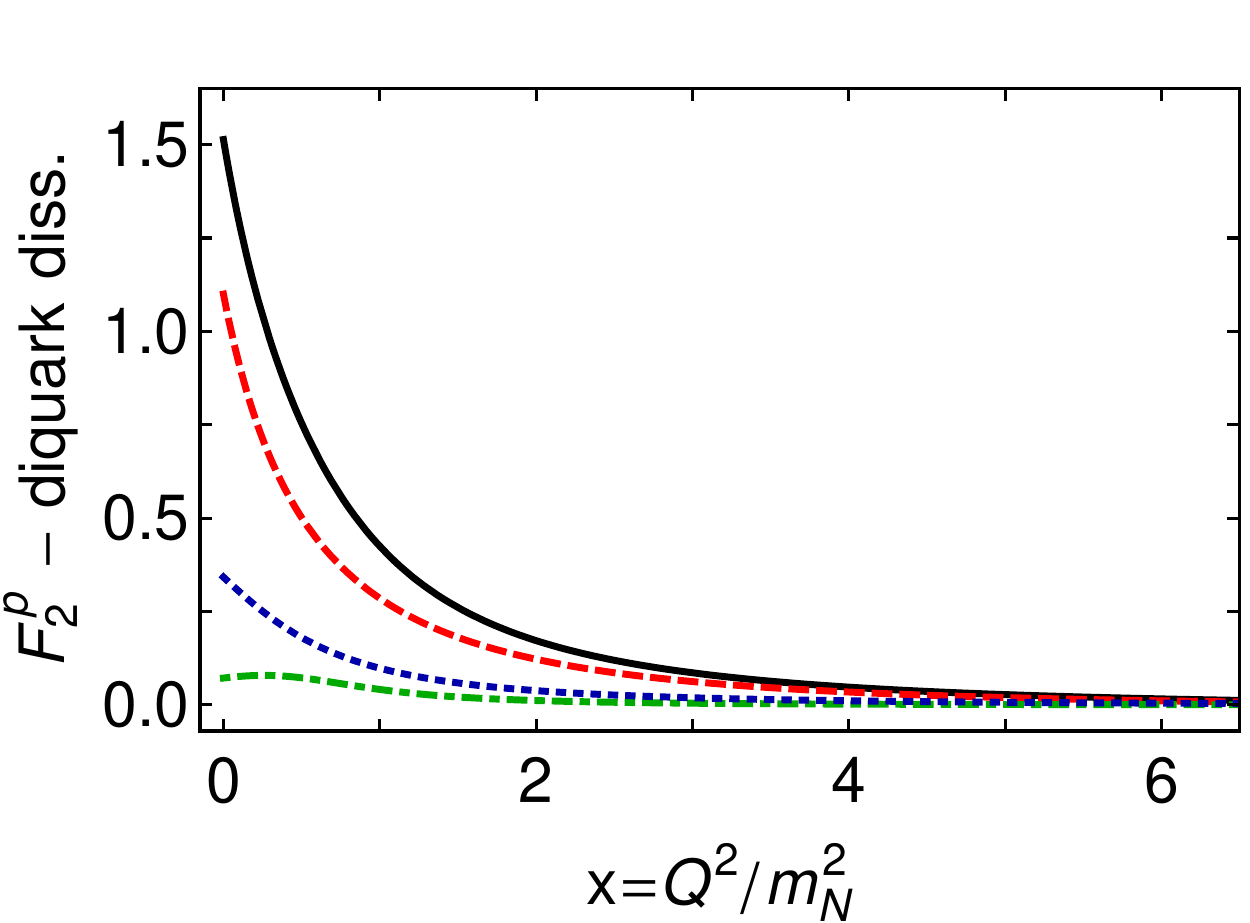}&
\includegraphics[clip,width=0.38\linewidth]{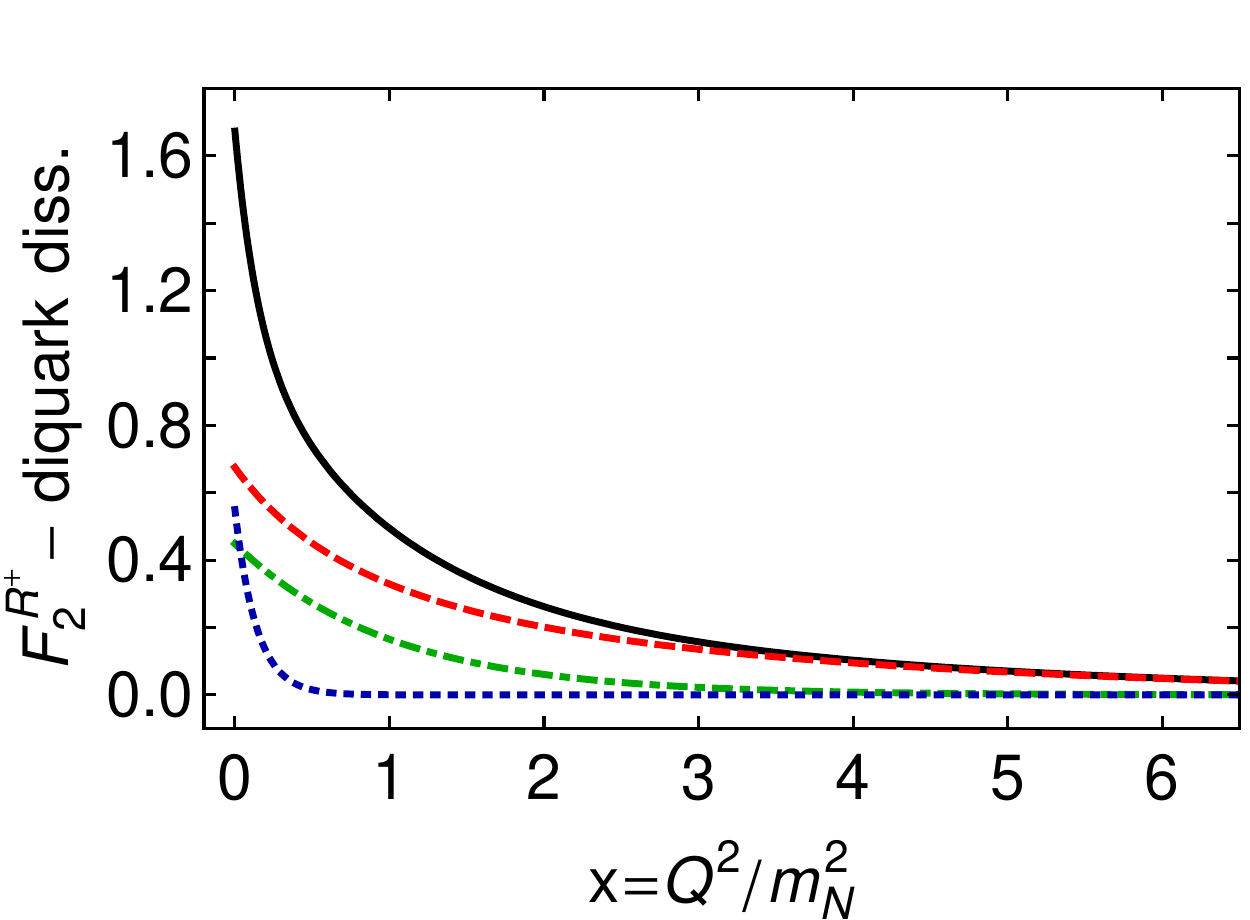}
\vspace*{-1.05cm}
\\
\vspace*{-0.5cm}
%
\includegraphics[clip,width=0.38\linewidth]{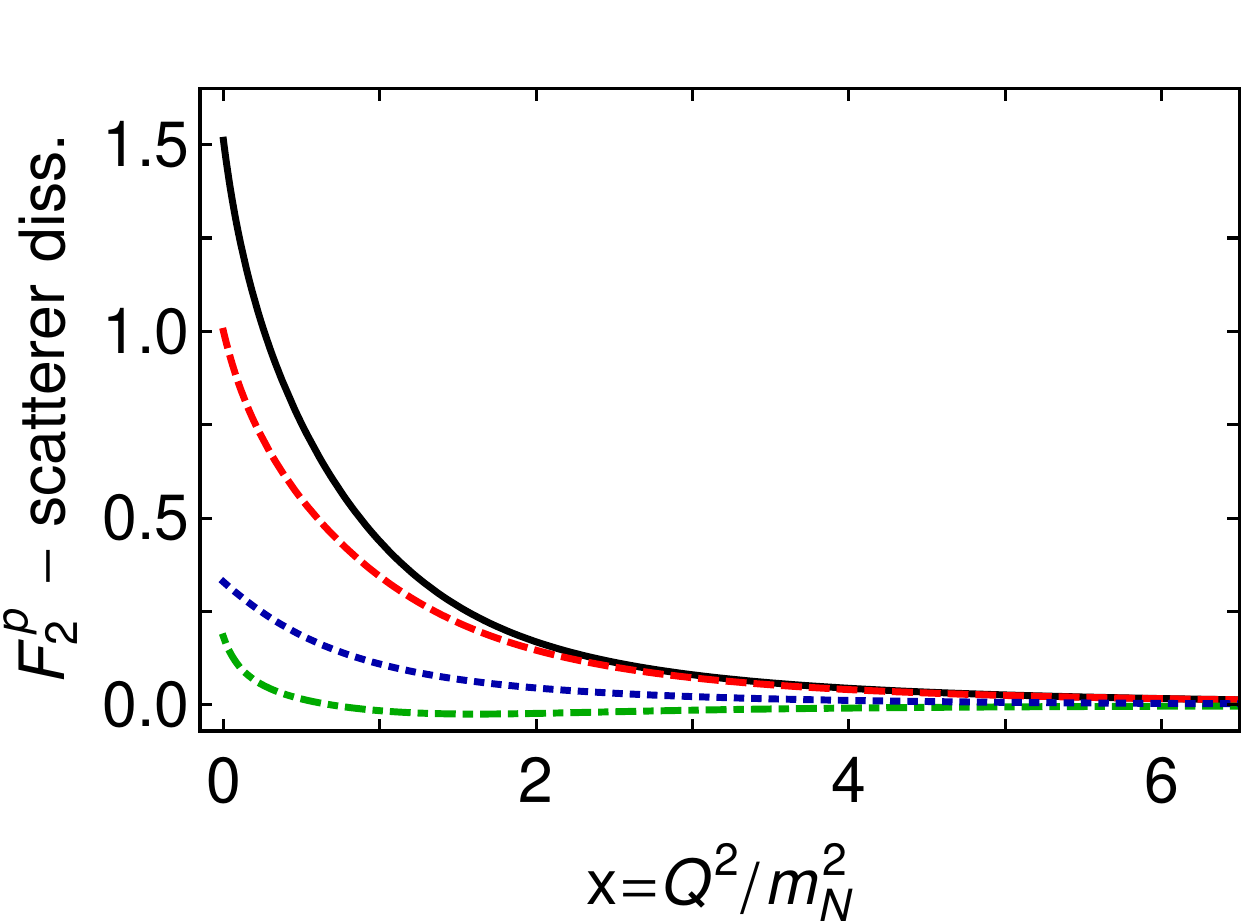} &
\includegraphics[clip,width=0.38\linewidth]{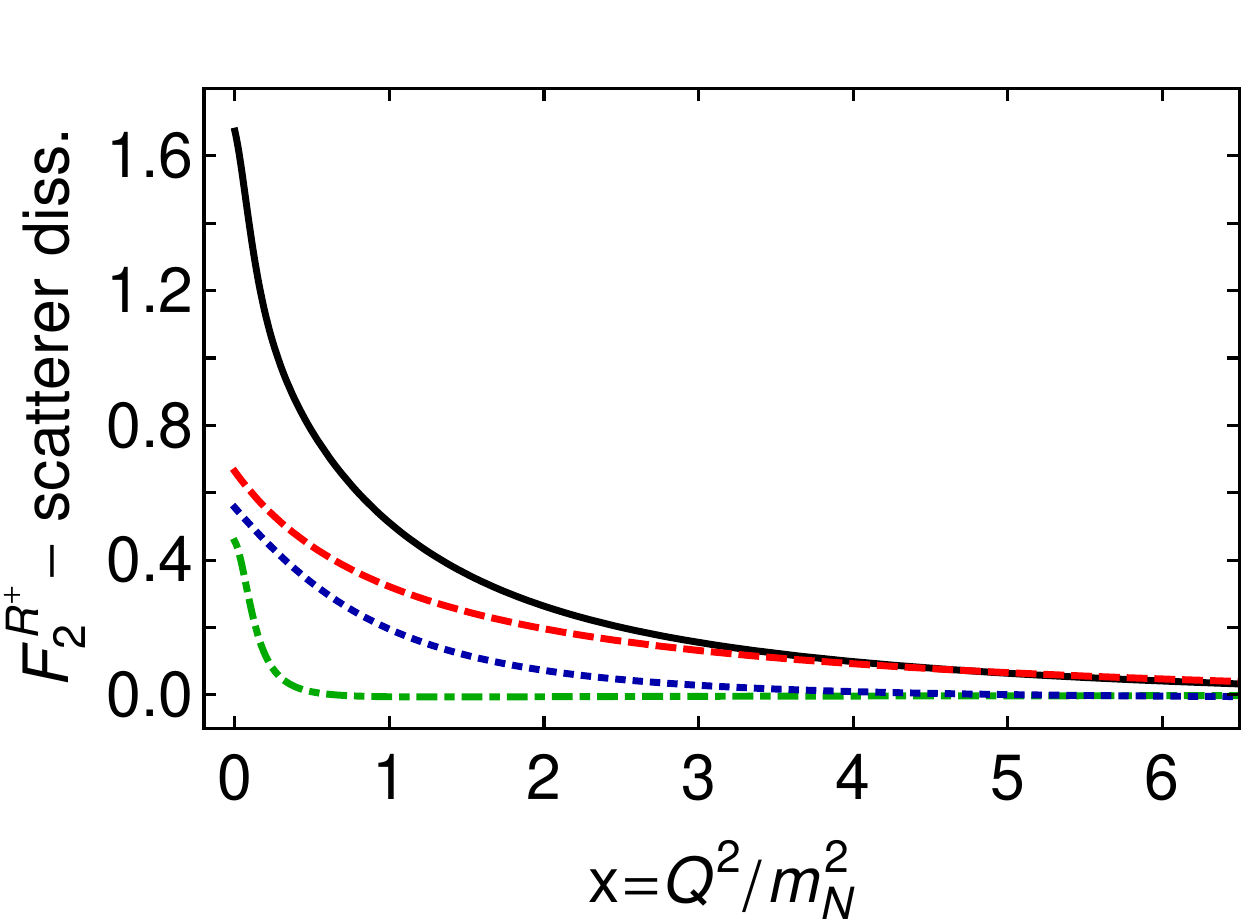}
\end{tabular}
\end{center}
\caption{\label{elasticdissectF2}
Pauli form factors of the proton and charged-Roper.
\emph{Upper panels} -- diquark breakdown: \emph{DD1} (dashed red), scalar diquark in initial and final baryon; \emph{DD2} (dot-dashed green), pseudovector diquark in both initial and final states; \emph{DD3} (dotted blue), scalar diquark in incoming baryon, pseudovector diquark in outgoing baryon, and vice versa.
\emph{Lower panels} -- scatterer breakdown: \emph{DS1} (red dashed), photon strikes an uncorrelated dressed quark; \emph{DS2} (dot-dashed green), photon strikes a diquark; and \emph{DS3} (dotted blue), diquark breakup contributions, including photon striking exchanged dressed-quark.
}
\end{figure*}

\begin{figure*}[!t]
\begin{center}
\begin{tabular}{lr}
\includegraphics[clip,width=0.38\linewidth]{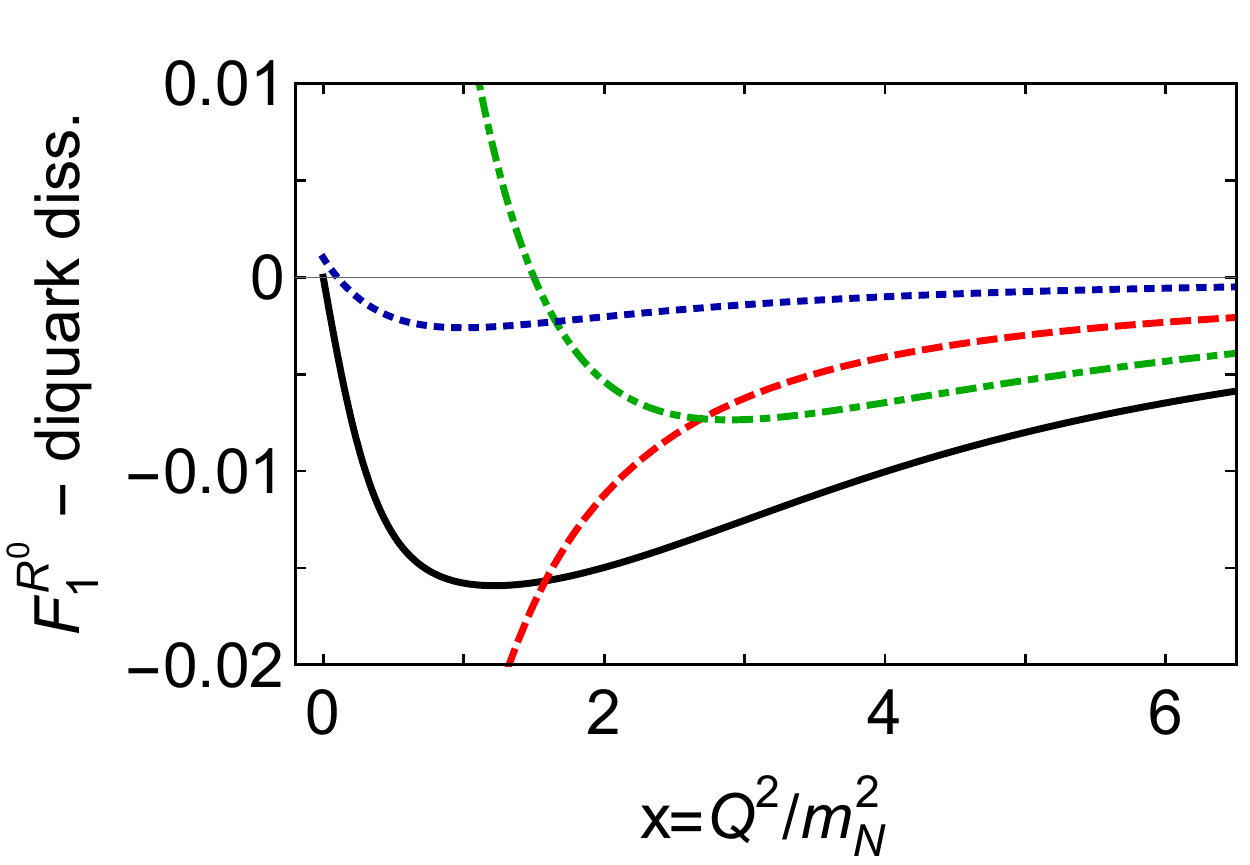} &
\includegraphics[clip,width=0.38\linewidth]{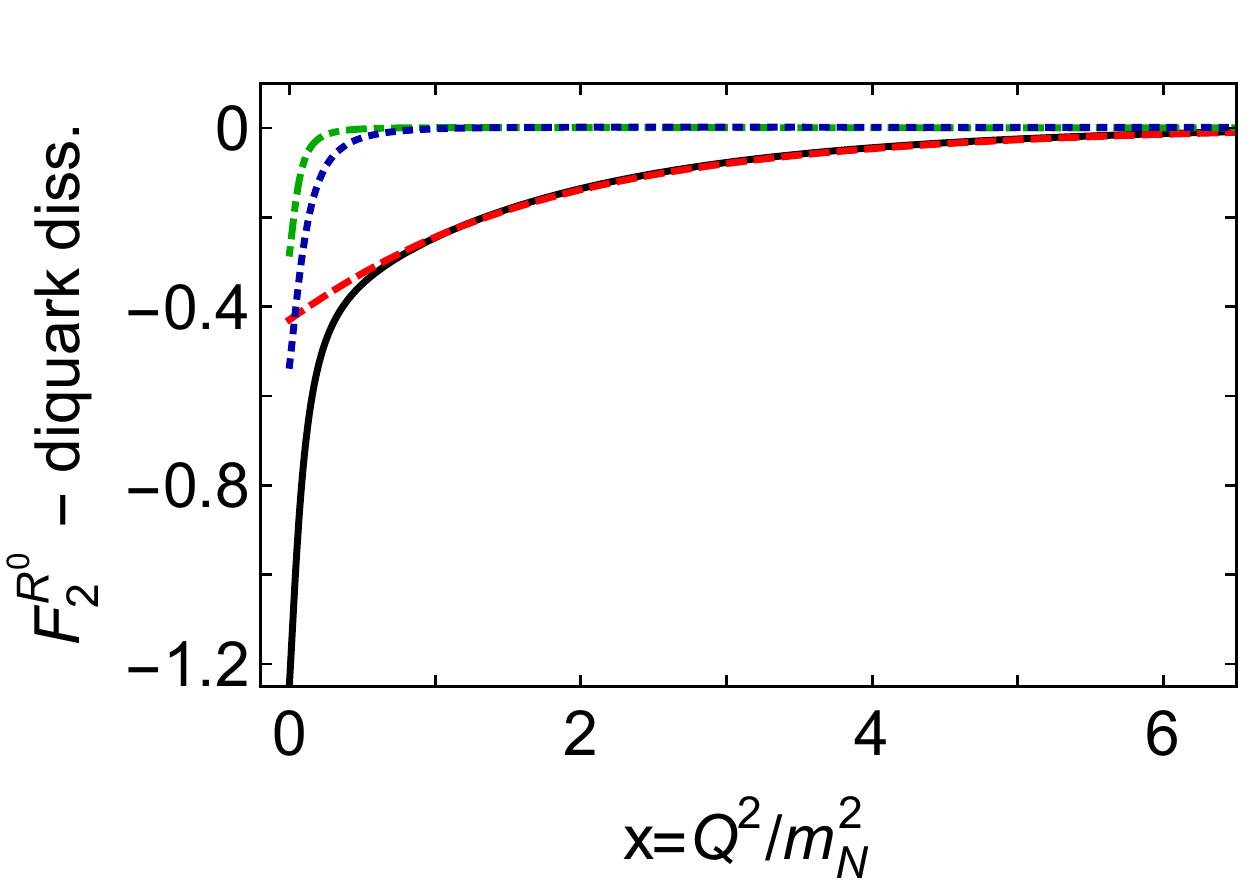}
\vspace*{-0.35cm}
\\
\vspace*{-0.35cm}
%
\includegraphics[clip,width=0.38\linewidth]{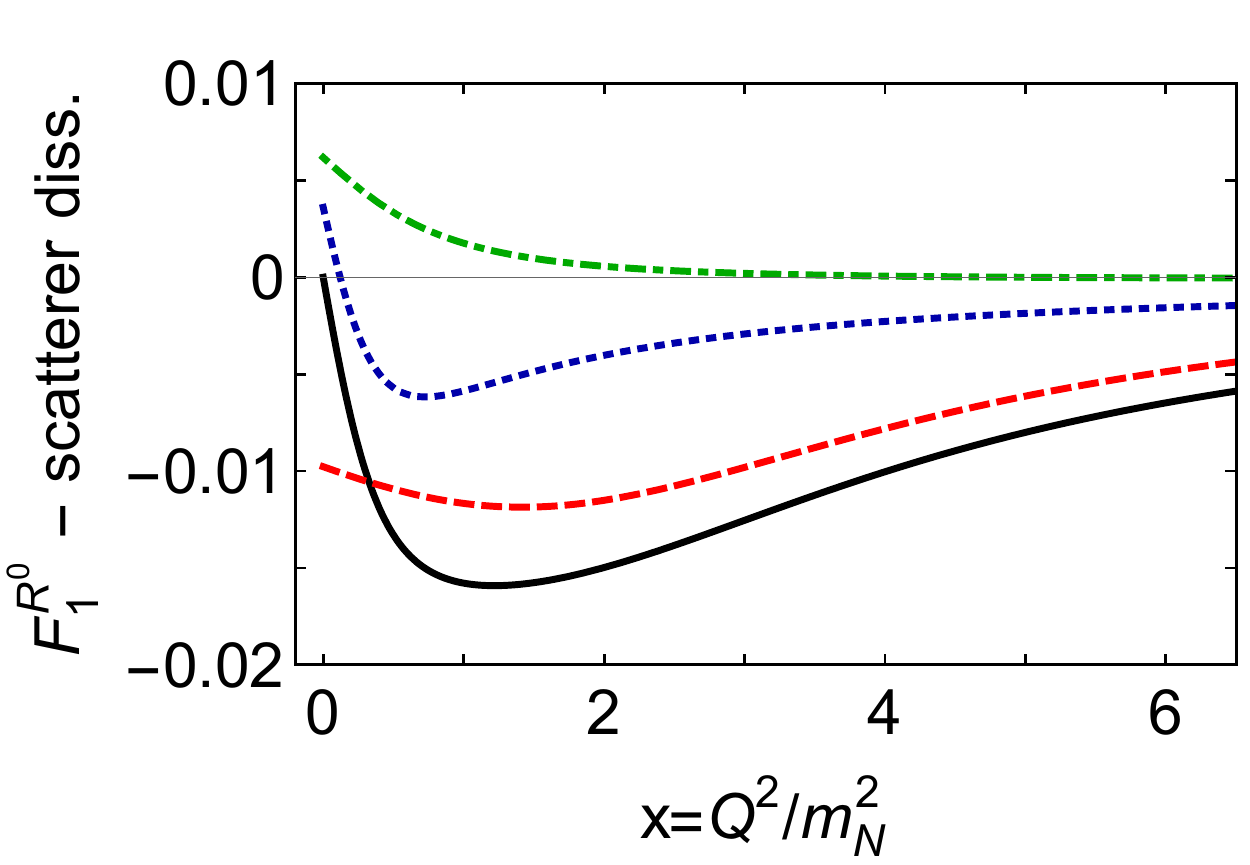} &
\includegraphics[clip,width=0.38\linewidth]{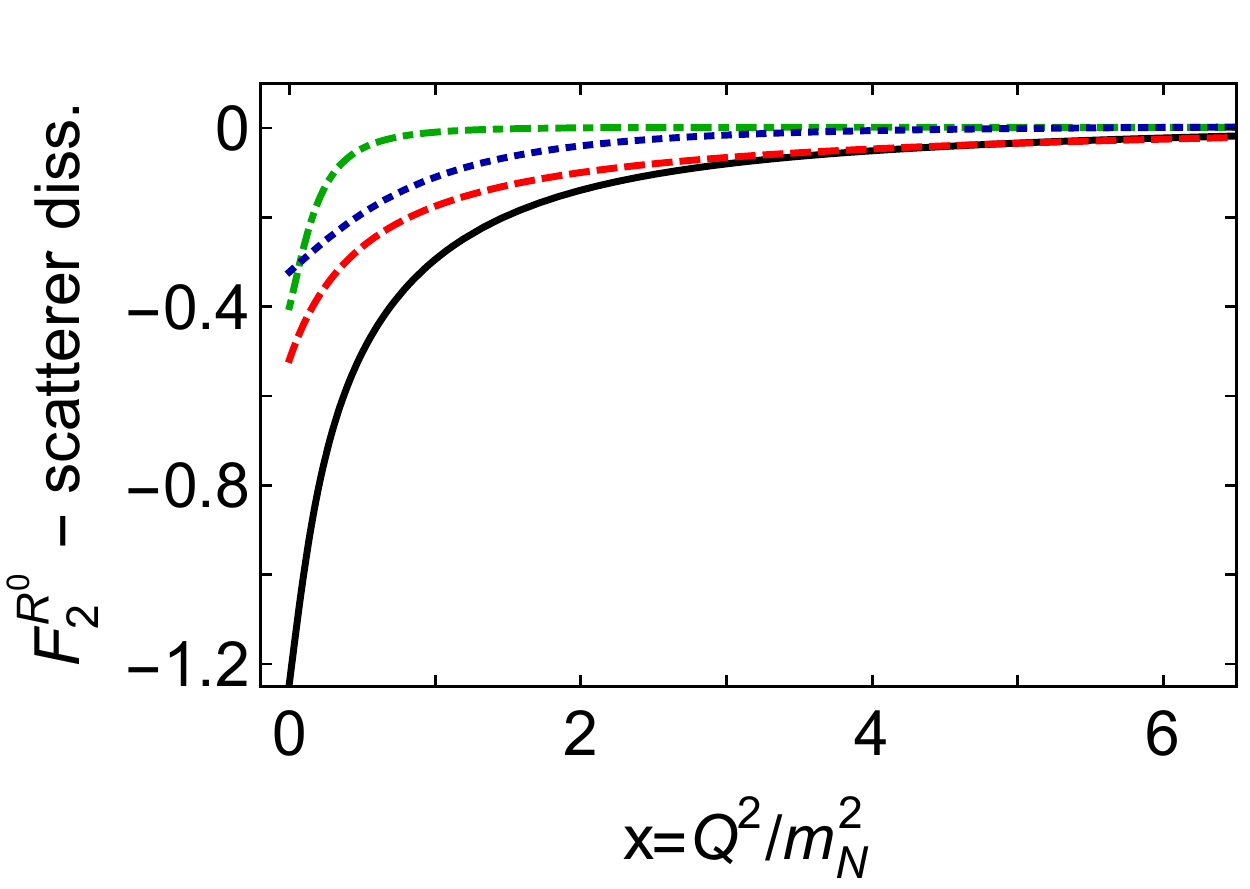}
\end{tabular}
\end{center}
\caption{\label{elasticdissectF12neutral}
Dirac (left) and Pauli (right) elastic form factors of neutral-Roper.
\emph{Upper panels} -- diquark breakdown: \emph{DD1} (dashed red), scalar diquark in initial and final baryon; \emph{DD2} (dot-dashed green), pseudovector diquark in both initial and final states; \emph{DD3} (dotted blue), scalar diquark in incoming baryon, pseudovector diquark in outgoing baryon, and vice versa.
\emph{Lower panels} -- scatterer breakdown: \emph{DS1} (red dashed), photon strikes an uncorrelated dressed quark; \emph{DS2} (dot-dashed green), photon strikes a diquark; and \emph{DS3} (dotted blue), diquark breakup contributions, including photon striking exchanged dressed-quark.
}
\end{figure*}

\section{Elastic Form Factors}
\label{secEMelastic}
The analogous elastic form factors for the states involved, $F_1^{f=i}(Q^2=0)$, are needed to fix the normalisation of the transition. The elastic Dirac and Pauli form factors associated with the dressed-quark core of the charged and neutral Roper are depicted in Fig.\,\ref{elastic} and compared with those for the proton and neutron from Ref.\,\cite{Segovia:2014aza}.  Evidently, there are qualitative similarities and quantitative differences, discussed at length in Ref.\,\cite{Chen:2018nsg}.

\begin{table}[b]
\caption{\label{tabstatic}
Static properties derived from the elastic form factors depicted in Fig.\,\ref{elastic}, see Eq.\,\eqref{eqradii} and following text.
($M_N = 1.18\,$GeV is the nucleon dressed-quark core mass.)}
\begin{center}
\begin{tabular*}
{\hsize}
{
l|@{\extracolsep{0ptplus1fil}}
c@{\extracolsep{0ptplus1fil}}
c@{\extracolsep{0ptplus1fil}}
c@{\extracolsep{0ptplus1fil}}
c@{\extracolsep{0ptplus1fil}}}\hline
  & $R^+$ & $p$ & $R^0$ & $n$ \\\hline
$r_E \, M_N$ & 6.23 & 3.65 & $\phantom{-}0.93i$ & $\phantom{-}1.67i$\\
$r_M \, M_N$ & 4.49 & 3.17 & $\phantom{-}4.15\phantom{i}$ & $\phantom{-}4.19\phantom{i}$\\
$\mu$ & 2.67 & 2.50 & $-1.24\phantom{i}$ & $-1.83\phantom{i}$ \\\hline
\end{tabular*}
\end{center}
\end{table}

Defining (Sachs) electric and magnetic form factors:
\begin{align}
G_E = F_1 - \frac{Q^2}{4 m_B^2} F_2\,,\quad
G_M = F_1 + F_2\,,
\end{align}
where $m_B$ is the baryon's mass, the $Q^2=0$ values and slopes of the form factors in Fig.\,\ref{elastic} yield the static properties listed in Table\,\ref{tabstatic}, where the radii are defined via
\begin{align}
\label{eqradii}
r^2 & = - \left.\frac{6}{\mathpzc n} \frac{d}{ dQ^2} G(Q^2)\right|_{Q^2=0}\,,
\end{align}
with ${\mathpzc n} = G(Q^2=0)$ when this quantity is nonzero, ${\mathpzc n} = 1$ otherwise, and the anomalous magnetic moment $\mu = G_M(0)$.
The electromagnetic radii of the charged-Roper core are larger than those of the proton core, but the magnetic moments are similar; and this pattern is reversed in the neutral-Roper/neutron comparison.

Figs.\,\ref{elasticdissectF1}, \ref{elasticdissectF2} display the contrast of diquark and scatterer dissections of the Dirac and Pauli form factors for
the proton and $R^+$. It is apparent from Fig.\,\ref{elasticdissectF1} that every contribution to the $R^+$ elastic Dirac form factor falls more rapidly than its analogue in the proton; while, on the other hand, the relative importance of each is typically the same within the proton and $R^+$. The former observation highlights that the relative strengths of the various diquarks in both the nucleon and Roper are almost identical. 
Concerning Fig.\,\ref{elasticdissectF2}, DD1$\times$DS1 is dominant in both columns, implying that the primary contribution to both proton and charged Roper comes from a photon striking a bystander quark in association with a $[ud]$-diquark; while differences appear for the subleading terms.  

Diquark and scatterer dissections for the Dirac and Pauli form factors of the neutral-Roper, which can be contrasted to $\gamma^\ast n \to R^0$ transition form factors, appear also displayed in Fig.\,\ref{elasticdissectF12neutral}.

\section{Nucleon-to-Roper Transition}
\label{secEMtransition}
\begin{figure*}[!t]
\begin{center}
\begin{tabular}{cc}
\includegraphics[clip,width=0.38\linewidth]{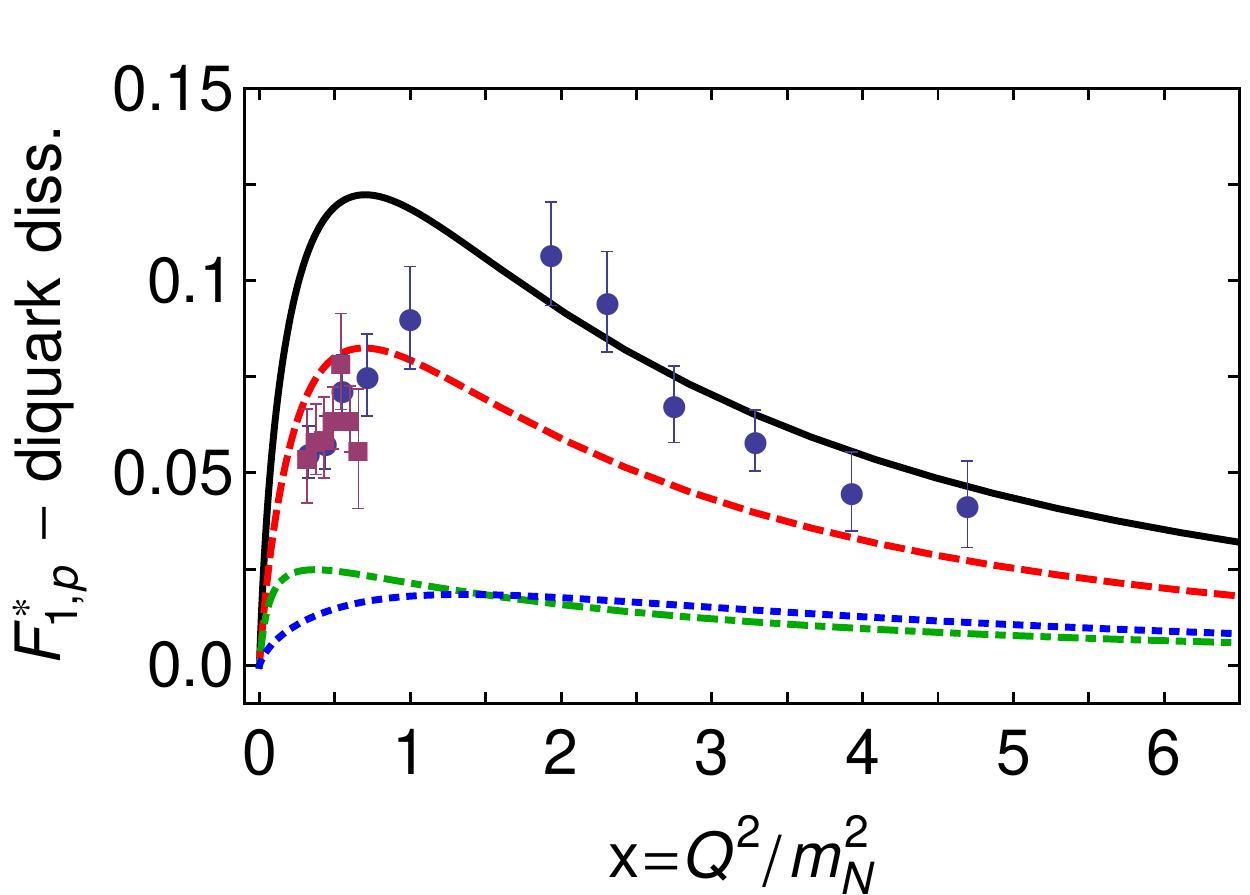} &
\includegraphics[clip,width=0.38\linewidth]{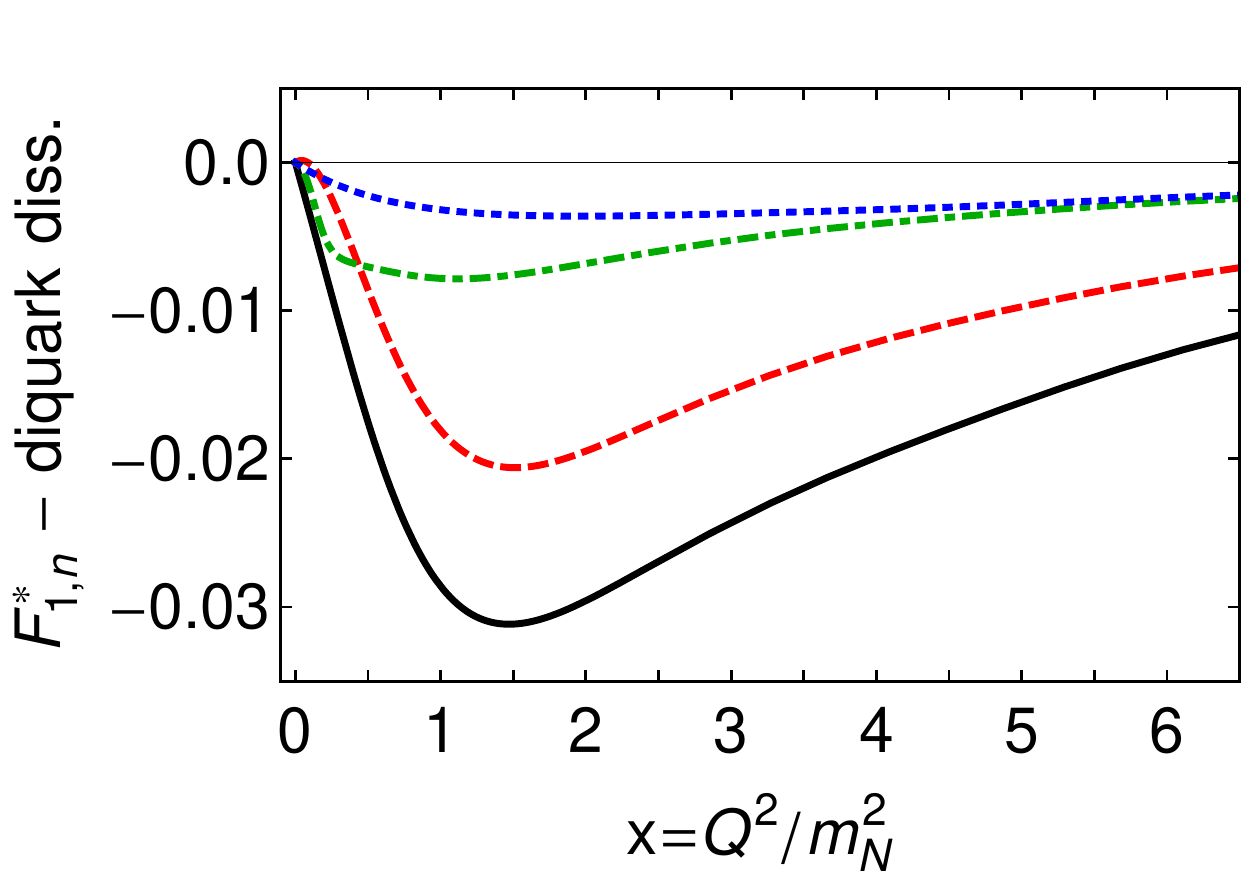}
\vspace*{-1.07cm}
\\
\vspace*{-0.5cm}
%
\includegraphics[clip,width=0.38\linewidth]{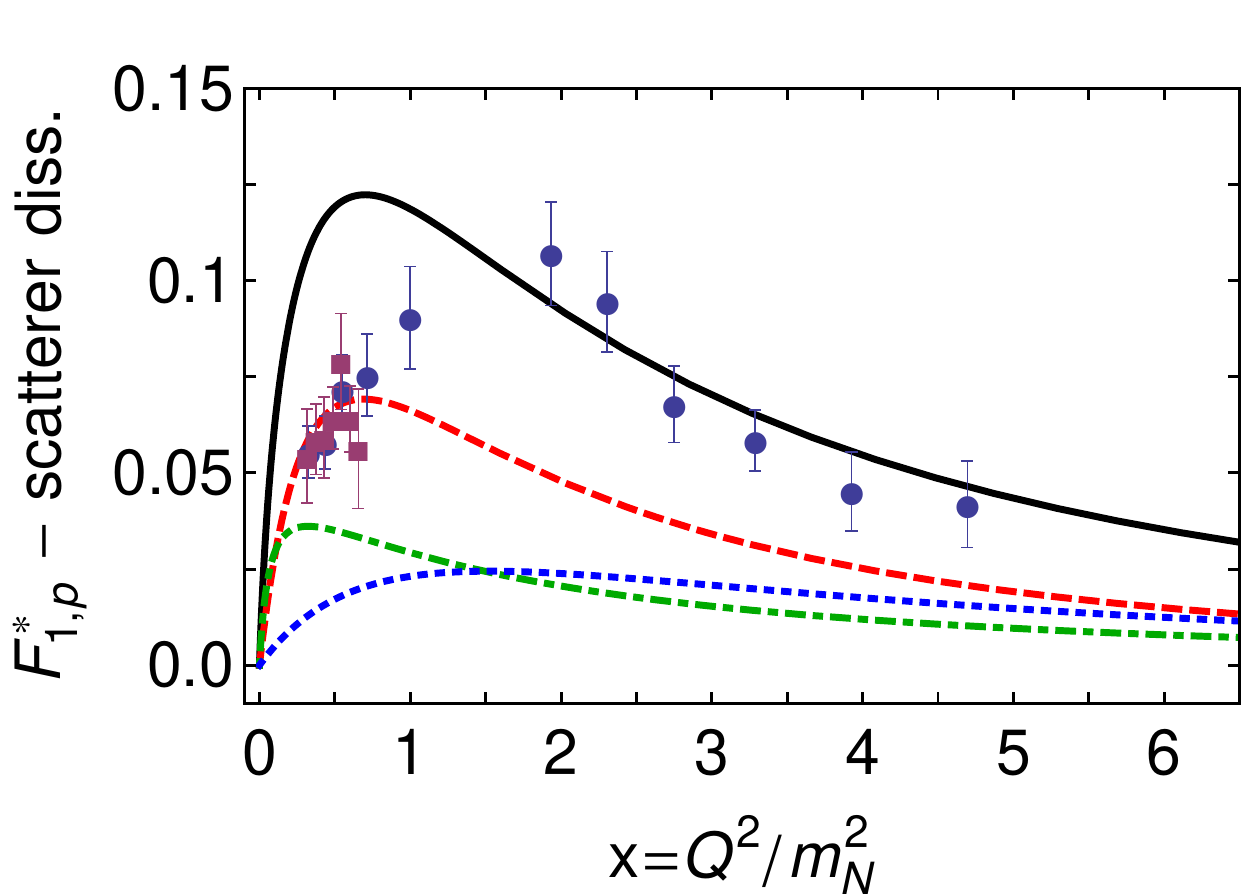} &
\includegraphics[clip,width=0.38\linewidth]{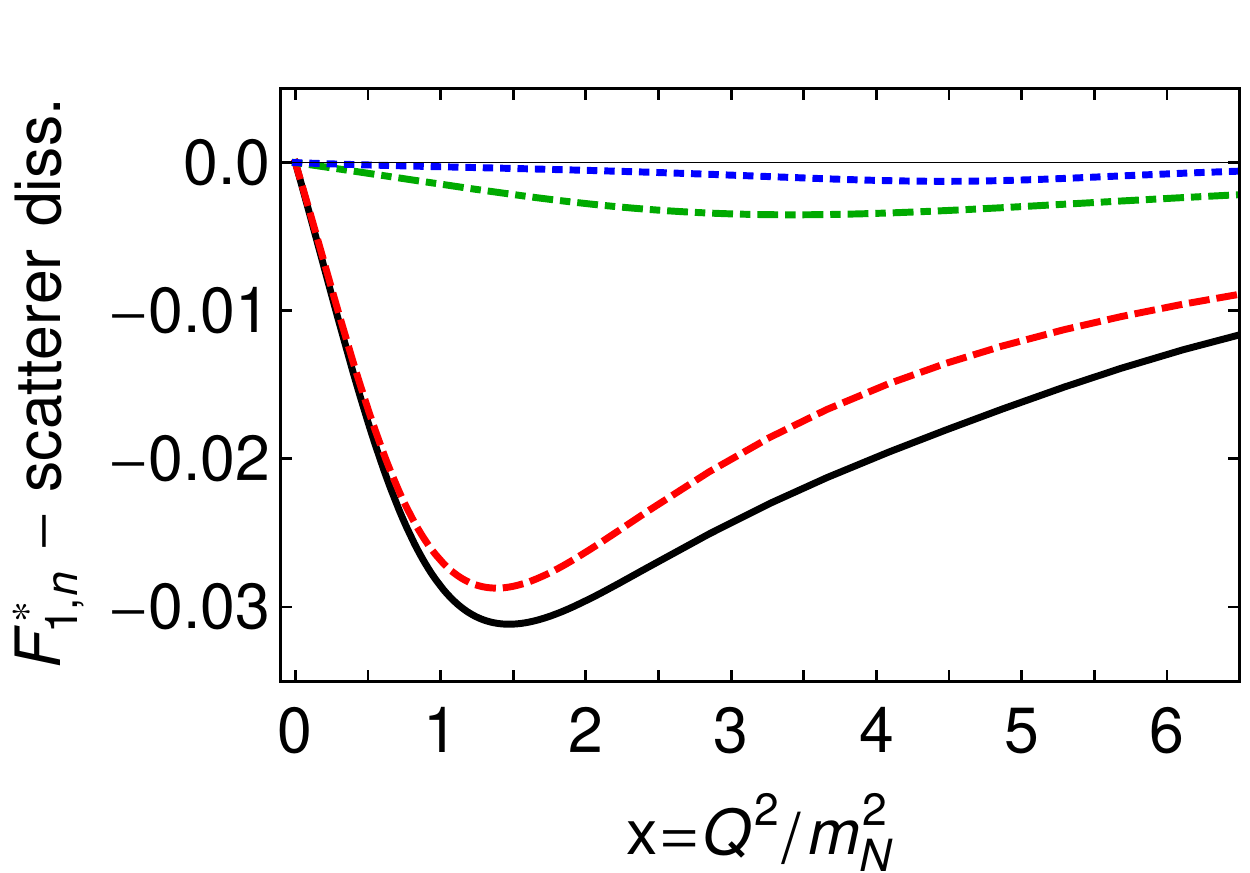}
\end{tabular}
\end{center}
\caption{\label{figF1}
Computed Dirac transition form factor, $F_{1}^{\ast}$, for the charged reaction $\gamma^\ast\,p\to R^+$ (left panels) and the neutral reaction $\gamma^\ast\,n\to R^0$ (right panels): solid (black) curve in each panel.
Data, left panels:
circles (blue)~\cite{Aznauryan:2009mx},
and squares (purple)~\cite{Mokeev:2012vsa, Mokeev:2015lda}.
\emph{Upper panels} -- diquark breakdown: \emph{DD1} (dashed red), scalar diquark in both nucleon and Roper; \emph{DD2}
(dot-dashed green), pseudovector diquark in both nucleon and Roper; \emph{DD3} (dotted blue), scalar diquark in
nucleon, pseudovector diquark in Roper, and vice versa.
\emph{Lower panels} -- scatterer breakdown: \emph{DS1} (red dashed), photon strikes an uncorrelated dressed quark;
\emph{DS2} (dot-dashed green), photon strikes a diquark; and \emph{DS3} (dotted blue), diquark breakup contributions,
including photon striking exchanged dressed-quark.
}
\end{figure*}

The Dirac transition form factors for $\gamma^\ast N \to R$, drawn in Fig.\,\ref{figF1}, show qualitative similarities to the elastic Dirac form factors for the charged channels depicted in Fig.\,\ref{elasticdissectF1}, dominated by the contribution in which a photon strikes a bystander dressed-quark in association with a $[ud]$-diquark, with lesser but non-negligible contributions from other processes.  
The same process for a photon striking a bystander dressed-quark partenered by $[ud]$ primarily contributes, too, to the transition in the neutral channel. Comparisons with the the left panels in Fig.\,\ref{elasticdissectF12neutral}, for $R^0$ elastic Dirac form factor, are however less transparent because, both $F_1^{R^0}$ and $F_{1,n}^\ast$ vainishing at the origin, in the latter case each contribution does separately, owing to state othogonality, while in the former charge neutrality only ensures that all terms sum to zero\,\cite{Chen:2018nsg}. 


Owing to the QCD-derived momentum-dependence of the propagators and vertices employed in solving the bound-state and scattering 
problems, $F_{1,p}^\ast(x)$ agrees quantitatively in magnitude and trend with the data above $x \simeq 2$. Below this value, the mismatch between prediction and data can be plausibly attributed to MB\,FSIs, as described in Sec.\,5 of Ref.\,\cite{Roberts:2016dnb}.  (See also Ref.\,\cite{Kamano:2018sfb}.)

A computation of the associated light-front-transverse transition charge-density is \cite{Tiator:2008kd}, 
\begin{align}
\label{eqrhob}
\rho^{pR}(|\vec{b}|)
& := \int \frac{d^2 \vec{q}_\perp }{(2\pi)^2} \,{\rm e}^{i \vec{q}_\perp \cdot \vec{b}} F_1^\ast(|\vec{q}_\perp|^2)\,,
%
\end{align}
the frame defined by $Q=(\vec{q}_\perp=(Q_1,Q_2),Q_3=0,Q_4=0)$, may be found elsewhere \cite{Roberts:2018hpf}, along with a related analysis of the impact of MB\,FSIs, which have been found to introduce significant attraction, screening the long negative tail of the quark-core contribution to $\rho^{pR}(|\vec{b}|)$ (consistently with their role in reducing the nucleon and Roper quark-core masses) and thereby compressing the transition domain in transverse space. Concerning the neutral transition, owing to its being uniformly small,  $F_{1,n}^\ast(x)$ can be plausibly impacted by MB\,FSIs on a larger $Q^2$-domain, as it appears to be the case, \emph{e.g}.\ with the electric quadrupole form factor in the $\gamma^\ast N \to \Delta$ transition \cite{Segovia:2014aza}.  

\begin{figure*}[!t]
\begin{center}
\begin{tabular}{cc}
\includegraphics[clip,width=0.38\linewidth]{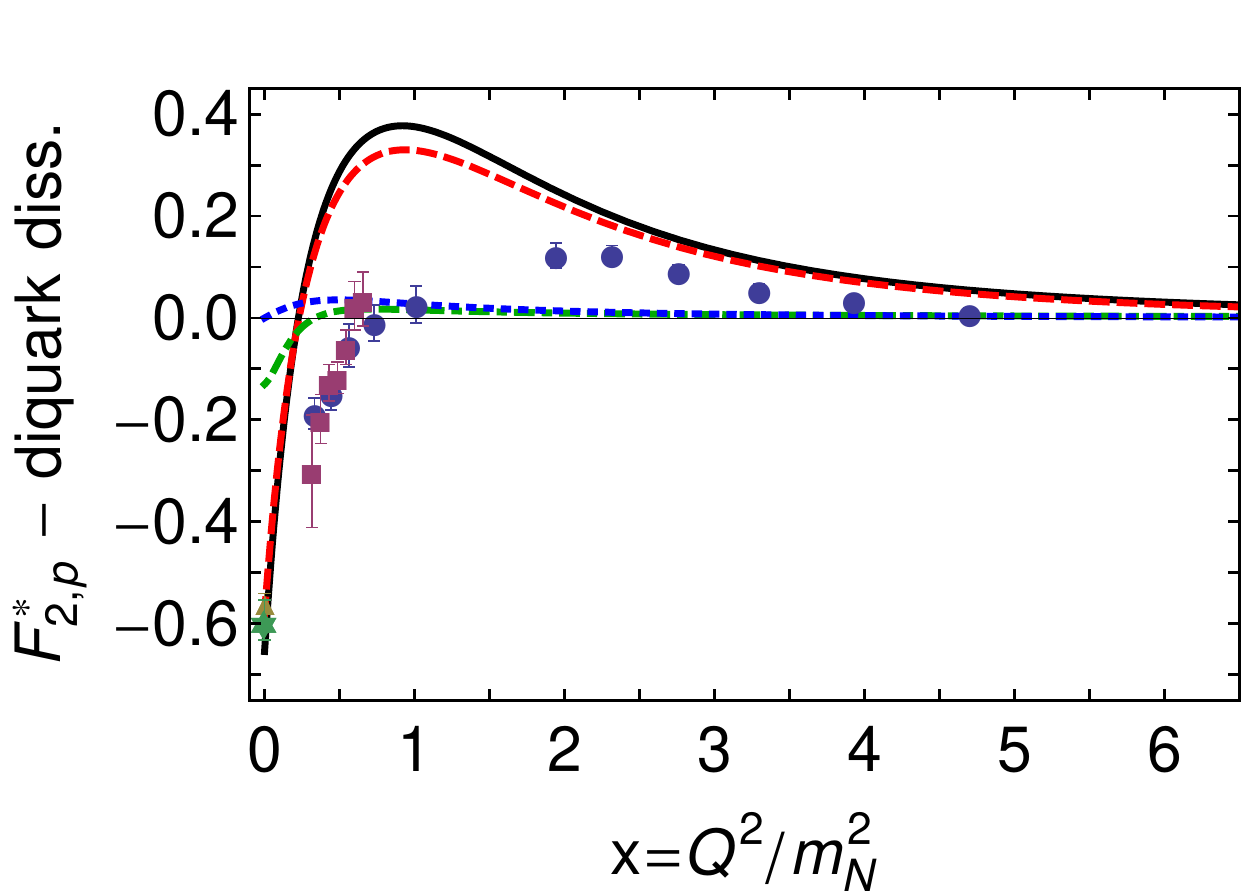} &
\includegraphics[clip,width=0.38\linewidth]{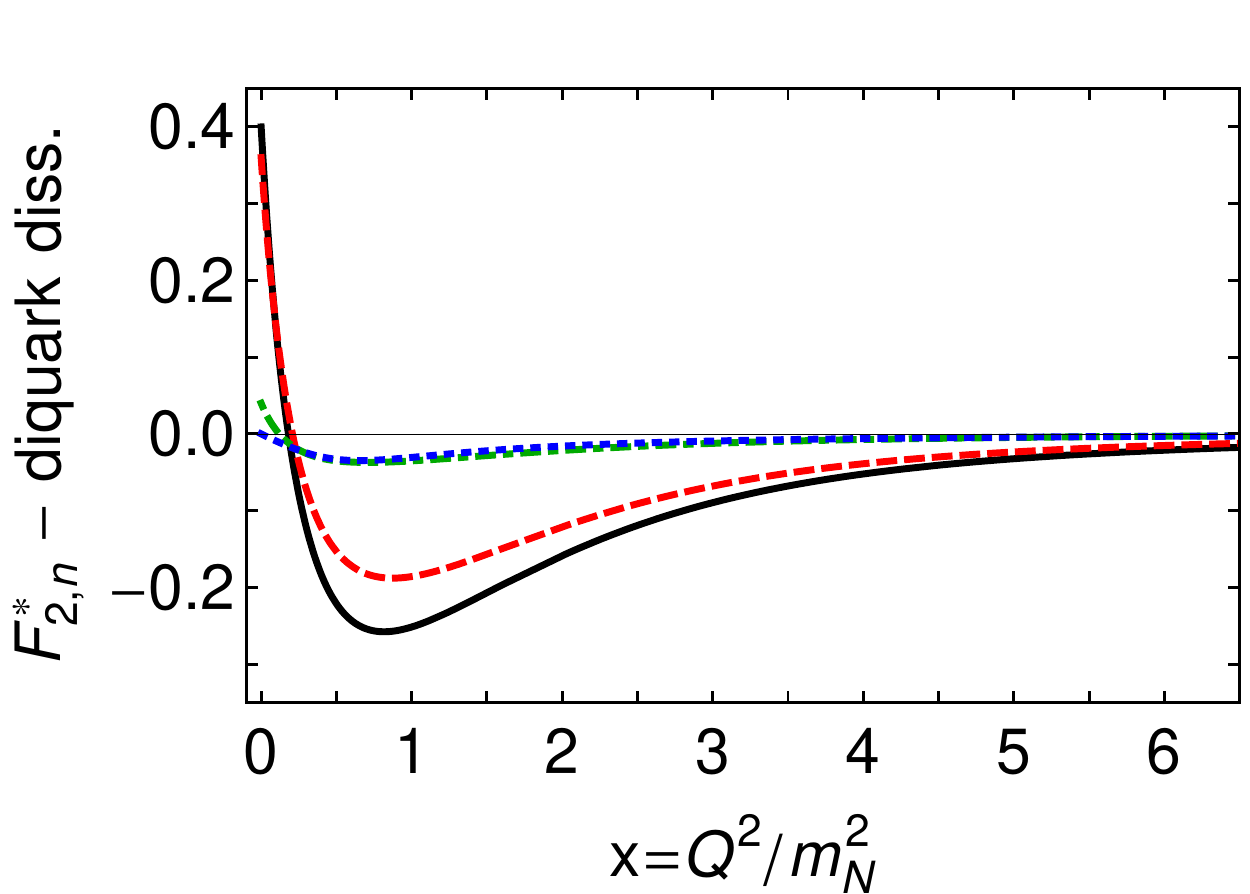}
\vspace*{-1.07cm}
\\
\vspace*{-0.5cm}
%
\includegraphics[clip,width=0.38\linewidth]{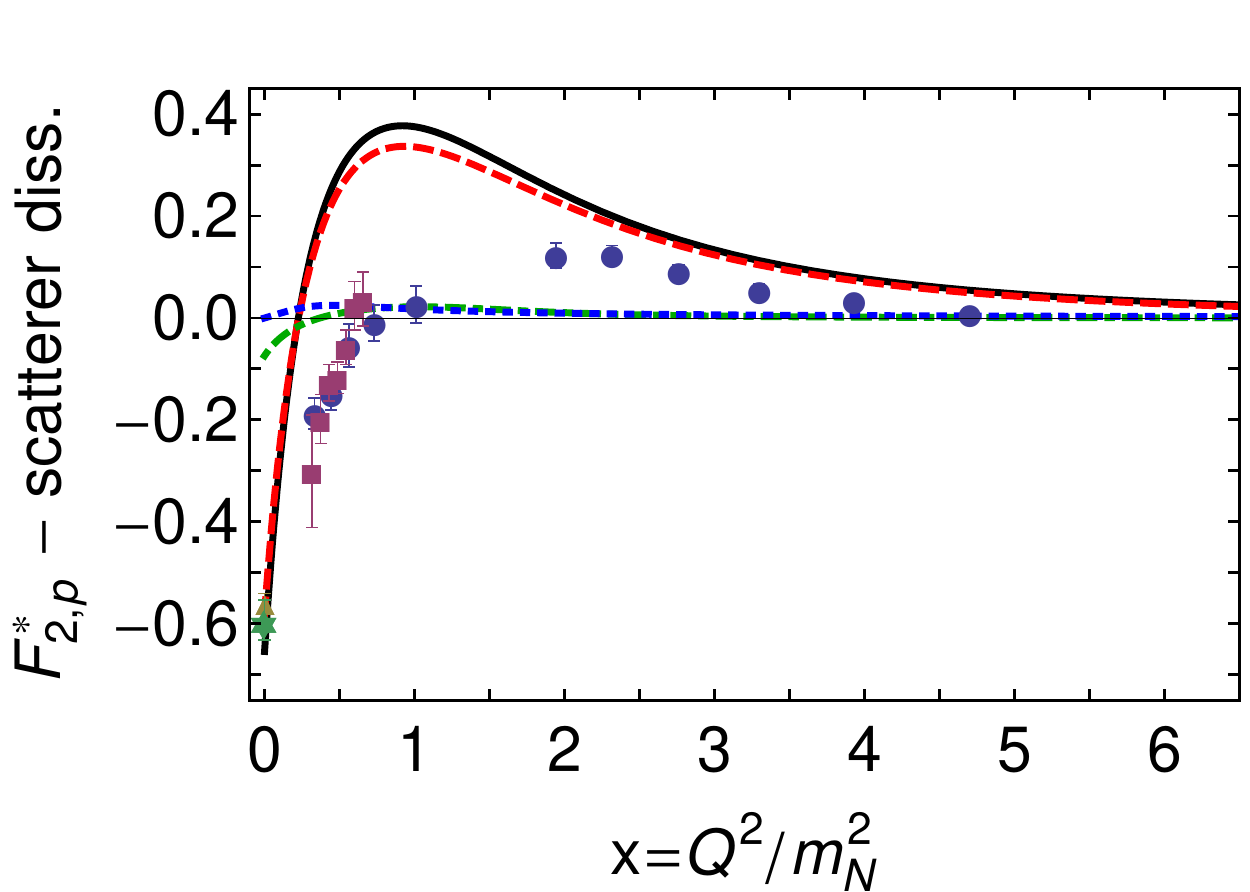}&
\includegraphics[clip,width=0.38\linewidth]{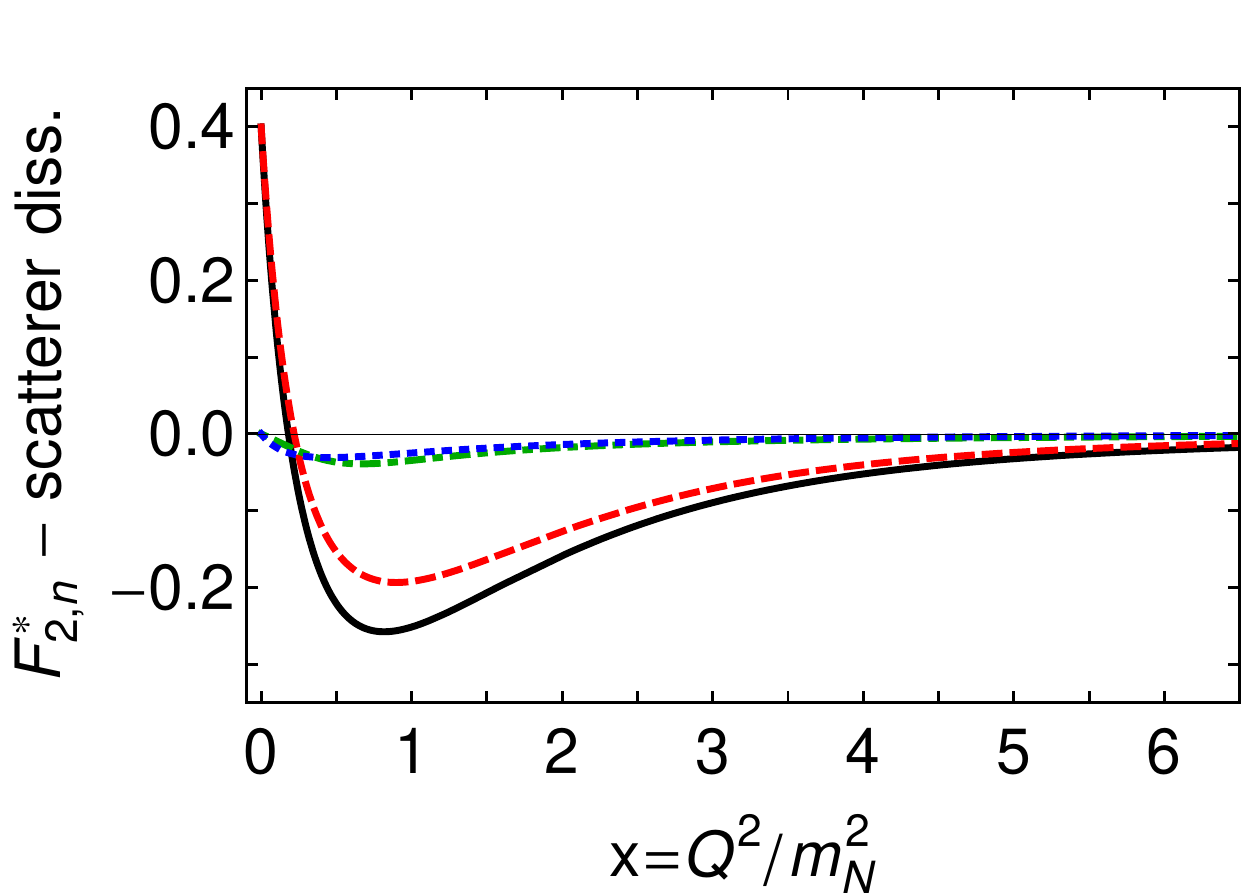}
\end{tabular}
\end{center}
\caption{\label{figF2}
Computed Pauli transition form factor, $F_{2}^{\ast}$, for the charged reaction $\gamma^\ast\,p\to R^+$ (left panels) and the neutral reaction $\gamma^\ast\,n\to R^0$ (right panels): solid (black) curve in each panel.
Data:
circles (blue)~\cite{Aznauryan:2009mx},
squares (purple)~\cite{Mokeev:2012vsa, Mokeev:2015lda},
triangle (gold)~\cite{Dugger:2009pn},
and star (green)~\cite{Tanabashi:2018oca}.
\emph{Upper panels} -- diquark breakdown: \emph{DD1} (dashed red), scalar diquark in both nucleon and Roper; \emph{DD2}
(dot-dashed green), pseudovector diquark in both nucleon and Roper; \emph{DD3} (dotted blue), scalar diquark in nucleon, pseudovector diquark in Roper, and vice versa.
\emph{Lower panels} -- scatterer breakdown: \emph{DS1} (red dashed), photon strikes an uncorrelated dressed quark;
\emph{DS2} (dot-dashed green), photon strikes a diquark; and \emph{DS3} (dotted blue), diquark breakup contributions,
including photon striking exchanged dressed-quark.
}
\end{figure*}

Pauli transition form factors for $\gamma^\ast N \to R$ are drawn in Fig.\,\ref{figF2}: They are all nonzero at $x=0$ and each crosses zero at roughly the same location, \emph{viz}.\, $x\approx 0.2$.  As happens with $F_2^{p,R^+}$ and $F_2^{n,R^0}$ in Fig.\,\ref{elastic}, $F_{2,p}^\ast$ and $F_{2,n}^\ast$ are similar in magnitude and $Q^2$-dependence and, remarkably, the ratio $F_{2,p}^\ast(0)/F_{2,n}^\ast(0) \approx -3/2$ is consistent with available data \cite{Tanabashi:2018oca}. Furthermore, MB\,FSIs also apply to Pauli form factors as to Dirac but, although affecting its precise location, do not spoil the existence of a zero in $F_{2}^{\ast}$, which can be confidently consisered as a robust prediction.   

\section{Transitions at Larger $Q^2$}
\label{SecLargeQ2}
We have recently reported\,\cite{Chen:2018nsg} projections for all transitions from factors on $x\in [0,12]$, aiming at anticipating the deliverance of data on the Roper-resonance electroproduction form factors out to $Q^2 \approx 12\,m_N^2$ in both the charged and neutral channels by the CLAS12 detector at JLab\,12. Let us sketch them here. 

Direct calculations of all the contributions to our transitions form factors in Fig.\,\ref{vertexB} imply eight-dimensional integrals to be evaluated (Diagrams~3, 5, 6), \emph{e.g.} by applying Monte-Carlo methods which are imprecise when they are required to deliver a small result and not all contributions share the same sign (it is the case with form factors at large photon virtuality). We then circumvented this difficulty in Ref.~\,\cite{Chen:2018nsg} by using the Schlessinger point method (SPM) \cite{Schlessinger:1966zz,Schlessinger:1968zz,Tripolt:2017pzb} to construct analytic approximations on $x\in [0,6]$, and then defining the results on $x \in [6,12]$ via the analytic continuation of those approximations.

\begin{figure}[!t] 
\includegraphics[clip,width=0.86\linewidth]{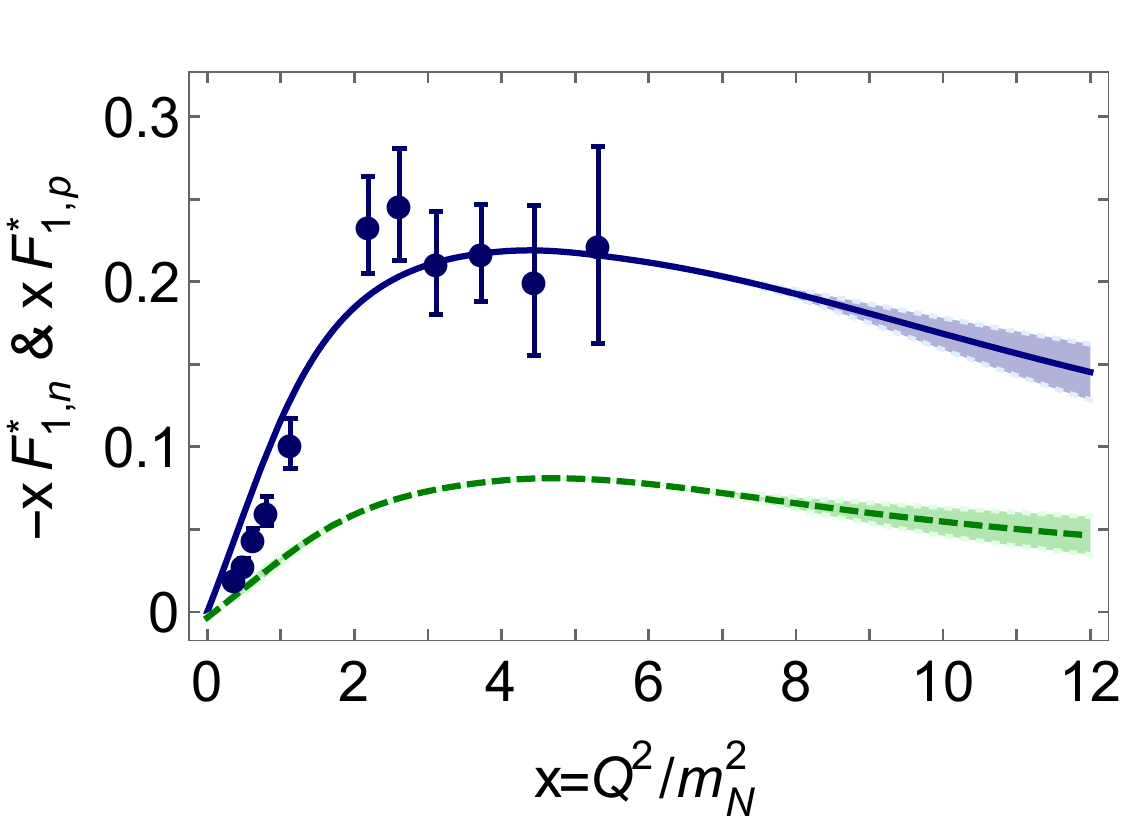} 
\includegraphics[clip,width=0.86\linewidth]{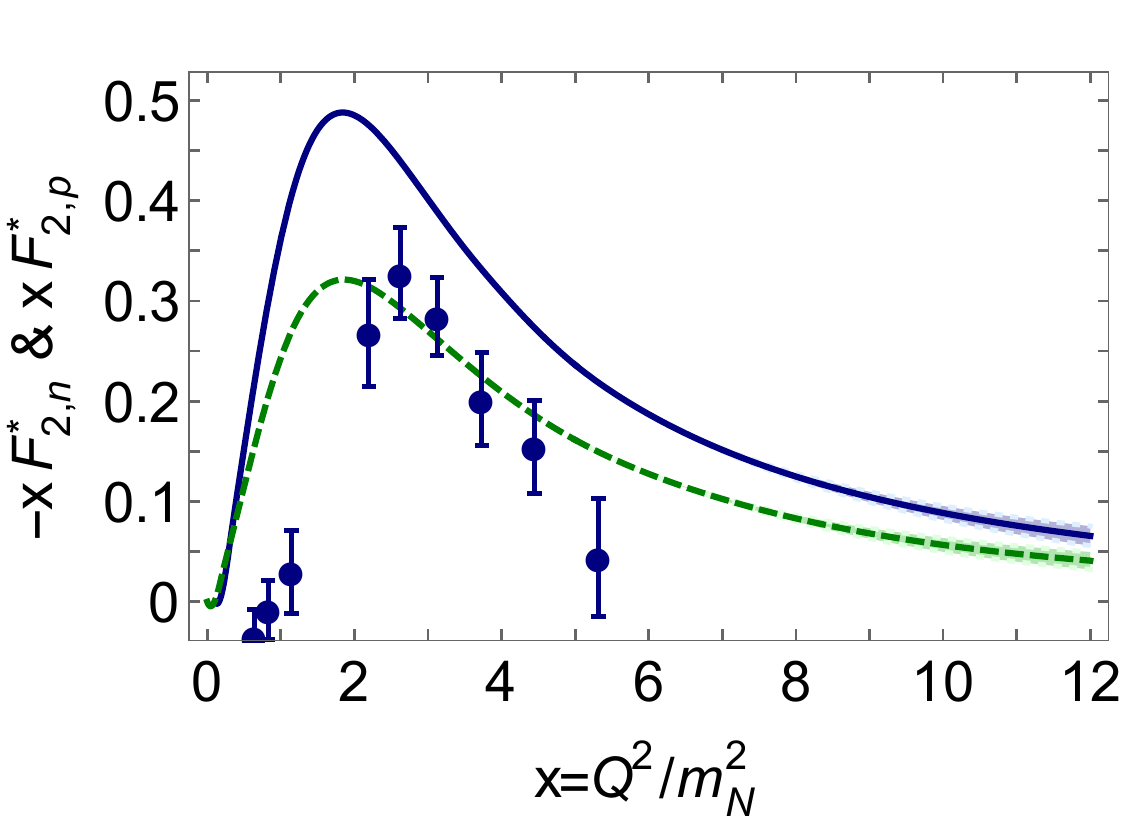}
\caption{\label{figLargeQ2}
Computed $x$-weighted Dirac (upper panel) and Pauli (lower panel) transition form factors for the reactions $\gamma^\ast\,p\to R^+$ (solid blue curves) and $\gamma^\ast\,n\to R^0$ (dashed green curves).  In all cases, the results on $x\in [6,12]$ are projections, obtained via extrapolation of analytic approximations to our results on $x\in [0,6]$: at each $x$, the width of the band associated with a given curve indicates our confidence in the extrapolated value.  (See text for details.)
Data in both panels are for the charged channel transitions, $F_{1,p}^\ast$ and $F_{2,p}^\ast$: circles (blue)~\cite{Aznauryan:2009mx}.  No data currently exist for the neutral channel.
}
\end{figure}

Details for the practical implementation of SPM in our case can be found in Ref.\,\cite{Chen:2018nsg}, suffice it to indicate here that we are eventually left with a band of extrapolated curves whose collective reliability at any $Q^2>Q_{\rm max}^2$ is expressed by the width of the band at that point, which is itself determined by the precision of the original output on $Q^2 \leq Q^2_{\rm max}$; with $Q^2_{\rm max}$ being the upper bound for the domain upon which direct computations have been confidently performed. Results for the $x$-weighted Dirac and Pauli transition form factors for the reactions $\gamma^\ast p \to R^{+}$, $\gamma^\ast n\to R^{0}$ appear displayed in Fig.\,\ref{figLargeQ2} on the domain $0<x<12$.  The precision of our SPM projections can be exemplified by quoting the form factor values at the upper bound of the extrapolation domain, $x_{12}=12$:
\begin{subequations}
\begin{align}
 F^\ast_{1,p} & = 0.0121(14) \,, \;  & -F^\ast_{1,n} & = 0.0039(10) \,,\\
x_{12}F^\ast_{1,p} & = 0.145(17) \,, \; & -x_{12}F^\ast_{1,n}  &= 0.046(11) \,,
\end{align}
\end{subequations}
\vspace*{-7ex}

\begin{subequations}
\begin{align}
F^\ast_{2,p} & = 0.0055(8) \,, \; & -F^\ast_{2,n}  &= 0.0034(7) \,,\\
x_{12}F^\ast_{2,p} & = 0.066(10) \,, \; & -x_{12}F^\ast_{2,n}  &= 0.041(9) \,.
\end{align}
\end{subequations}

The $x$-weighted results accentuate, without overmagnifying, the larger-$x$ behaviour of the form factors.  On the domain depicted, there is no indication of the scaling behaviour expected of the transition form factors: $F^\ast_{1} \sim 1/x^2$, $F^\ast_2 \sim 1/x^3$. One would anyhow expect such behaviour becoming manifest on $x\gtrsim 20$, since each dressed-quark in the baryons must roughly share the impulse momentum, $Q$.

\section{Conclusions}
\label{secEpilogue}
The existence of strong nonpointlike quark-quark (diquark) correlations within baryons has long been argued. On the ground of this, we used a quark-diquark approximation to the Poincar\'e-covariant three-body bound-state problem to compute all form factors relevant to the $\gamma^\ast N \to R$ transitions in Ref.\,\cite{Chen:2018nsg} and have now reported here the results. We have found that both scalar and pseudovector diquarks are essential for a description of existing data, but correlations in other diquark channels can be neglected. The same holds true for nucleon elastic and $\gamma^\ast N \to \Delta(1232)$ transition form factors.

Focusing on $\gamma^\ast N \to R$, precise measurements in the charged channel already exist 
\cite{Aznauryan:2009mx, Aznauryan:2012ec, Aznauryan:2012ba, Park:2014yea,Isupov:2017lnd,Fedotov:2018oan}, novel experiments are approved at JLab\,12 and elsewhere, and others are either planned or under consideration as part of an international effort to measure transition electrocouplings of all prominent nucleon resonances \cite{Aznauryan:2012ba,Mokeev:2018zxt,Carman:2018fsn,Cole:2018faq,Ramstein:2018wkk}.  Hence, our predictions, including those in the neutral channel, are likely to be tested in the foreseeable future.

The empirical information that can be delivered by such experiments has the potential to address a wide range of issues. In particular, it should help to confirm whether, as happens for the nucleon, $\Delta$-baryon and Roper resonance, the quark-quark correlations play an essential role in the structure of all baryons. And, underlying this, one could be able to inquire about whether the expression of DCSB is the same in each baryon. Indeed, the nucleon, $\Delta$-baryon and Roper resonance are merely a small collection of closely-related positive-parity baryons; and, hence, consistency with available data may not be seen as conclusive. Even less conclusive because there can be emerging evidences indicating that pseudoscalar and vector diquark correlations also play a material role in low-lying negative-parity baryons \cite{Eichmann:2016hgl, Lu:2017cln, Chen:2017pse}, that excited states of the $\Delta$-baryon possess unexpectedly complicated wave functions \cite{Qin:2018dqp}, and very little is known about the Poincar\'e-covariant wave functions of $I=1/2$, $J=3/2$ baryons.  These systems are the focus of forthcoming analyses using the methods herein described.  

\section*{acknowledgements}

We are grateful to R. W. Gothe for convening the session in which this perspective was presented. Work partly supported by Spanish ministry project PPA2017-86380-P and by Jiangsu Province Hundred Talents Plan for Professionals.

\bibliography{refs}

\end{document}